 \documentclass[useAMS,usenatbib]{mnras}
\usepackage{amsmath}
\usepackage{appendix}
\usepackage{graphicx}
\usepackage{xfrac}
\usepackage{caption}
\usepackage{array}
\usepackage{longtable}
\usepackage{pdflscape}
\usepackage[figuresright]{rotating}
\usepackage{booktabs}
\usepackage[usenames, dvipsnames]{color}
\usepackage[flushleft]{threeparttable}
\usepackage[export]{adjustbox}
\usepackage{hyperref}

\begin{document}

\title[GAMA Panchromatic Photometry]{Galaxy And Mass Assembly (GAMA): Assimilation of KiDS into the GAMA database}

\author[Bellstedt et al.]{Sabine Bellstedt,$^1$\thanks{E-mail: sabine.bellstedt@uwa.edu.au} Simon P.~Driver,$^{1,2}$   Aaron S. G. Robotham,$^1$ \newauthor Luke J. M. Davies,$^1$ 
Cameron R. J. Bogue,$^{1,3}$ Robin H. W. Cook,$^1$ \newauthor Abdolhosein Hashemizadeh,$^1$  Soheil Koushan,$^1$ Edward N. Taylor,$^{4}$ \newauthor Jessica E. Thorne,$^1$  Ryan J. Turner,$^1$ Angus H. Wright$^{5,6}$ \\
$^1$ICRAR, The University of Western Australia, 35 Stirling Highway, Crawley WA 6009, Australia\\
$^2$SUPA, School of Physics \& Astronomy, University of St Andrews, North Haugh, St Andrews, KY16 9SS, UK\\
$^3$School of Physics and Astronomy, Cardiff University, Queens Buildings, The Parade, Cardiff CF24 3AA, UK\\
$^4$Centre for Astrophysics and Supercomputing, Swinburne University of Technology, Hawthorn VIC 3122, Australia\\
$^5$Astronomisches Institut, Ruhr-Universit{\"a}t Bochum, Universit{\"a}tsstr. 150, 44801 Bochum, Germany\\
$^6$Argelander-Institut f{\"u}r Astronomie, Auf dem H{\"u}gel 71, 53121 Bonn, Germany
}


\pubyear{2018}


\label{firstpage}
\pagerange{\pageref{firstpage}--\pageref{lastpage}}
\maketitle

\begin{abstract}
The Galaxy And Mass Assembly Survey (GAMA) covers five fields with highly complete spectroscopic coverage ($>95$ per cent) to intermediate depths ($r<19.8$ or $i < 19.0$ mag), and collectively spans 250 square degrees of Equatorial or Southern sky. 
Four of the GAMA fields (G09, G12, G15 and G23) reside in the ESO VST KiDS and ESO VISTA VIKING survey footprints, which combined with our GALEX, WISE and Herschel data provide deep uniform imaging in the $FUV\,NUV\,ugriZYJHK_s\,W1\,W2\,W3\,W4\,P100\,P160\,S250\,S350\,S500$ bands. 
Following the release of KiDS DR4, we describe the process by which we ingest the KiDS data into GAMA (replacing the SDSS data previously used for G09, G12 and G15), and redefine our core optical and near-IR catalogues to provide a complete and homogeneous dataset. 
The source extraction and analysis is based on the new \textsc{ProFound} image analysis package, providing matched-segment photometry across all bands. 
The data are classified into stars, galaxies, artefacts, and ambiguous objects, and objects are linked to the GAMA spectroscopic target catalogue. 
Additionally, a new technique is employed utilising \textsc{ProFound} to extract photometry in the unresolved MIR-FIR regime.
The catalogues including the full FUV-FIR photometry are described and will be fully available as part of GAMA DR4. 
They are intended for both standalone science, selection for targeted follow-up with 4MOST, as well as an accompaniment to the upcoming and ongoing radio arrays now studying the GAMA $23^h$ field.
\end{abstract}


\setlength{\extrarowheight}{0pt}
\section{Introduction}
The era of the modern wide-area imaging survey, i.e., those based on linear digital detectors and covering a sizeable portion of the sky, started in earnest with the 2MASS\footnote{Two Micron All-Sky Survey} \citep{Skrutskie06}, SDSS\footnote{Sloan Digital Sky Survey} \citep{York00}, and UKIDSS\footnote{UKIRT Infrared Deep Sky Survey} \citep{Lawrence07} surveys --- although it would be remiss not to mention the equally transformational IRAS\footnote{Infrared Astronomical Satellite} \citep{Neugebauer84} and ROSAT\footnote{ROentgen SATellite} \citep{Voges99} space missions. These programs, as well as achieving transformational science from the Solar System to the distant Universe, have in turn motivated the emergence of a multitude of dedicated imaging facilities on the ground including, for example, SkyMapper \citep{Keller07}; VST\footnote{VLT Survey Telescope} \citep{Arnaboldi07}; VISTA\footnote{Visible and Infrared Survey Telescope for Astronomy} \citep{Sutherland15}; and LSST\footnote{Large Synoptic Survey Telescope} \citep{Ivezic19}, and in space, for example GALEX\footnote{Galaxy Evolution Explorer} \citep{Martin05}; WISE\footnote{Wide-field Infrared Survey Explorer} \citep{Wright10}; Spitzer \citep{Werner04}; Herschel \citep{Pilbratt10}; Euclid \citep{Beaulieu10}; and WFIRST\footnote{Wide Field Infrared Survey Telescope} \citep{Gehrels15} to highlight a few. 

The further {\it federation} of these data streams with ground-based spectroscopy and other facilities, has allowed for the construction of a truly multi-wavelength and three-dimensional view of our Universe \citep[e.g.,][]{Jarrett17, Driver18}. In particular, major advances have been made in
quantifying: the spatial distribution of galaxies and their use for cosmology; the distribution of groups and clusters; studies of galaxy populations; galaxy merger rates; the assembly of mass (stellar, dust, gas and super-massive black holes); the transformation of mass; and identified the primary energy production pathways (star-formation and Active Nuclei); all as a function of look-back time and environment. The data and science from these surveys now dominate our knowledge of the near and intermediate Universe, and provide the vital zero redshift benchmark for studies of the distant and adolescent Universe. 

Not only has our knowledge and understanding been advanced, but also the way in which astronomy is conducted, shifting from an individual to team pursuit \citep{Milojevic14}. Collectively, these {\it major} endeavours have allowed us to start the process of {\it comprehensively} mapping the evolution of all mass, energy, and structure over all cosmic time and to build the scaffolding upon which the numerical N-body, hydrodynamic and semi-analytic models hang \citep[e.g. ][]{Lagos19}. In the coming years such comprehensive studies will be massively augmented with new wide-area optical/near-IR (LSST, Euclid, WFIRST), x-ray (eROSITA), and deep radio (MeerKAT, ASKAP, MWA and SKA) imaging and spectral surveys, taking us from a multi-wavelength outlook, to a truly panchromatic perspective. 

While acknowledging this impending paradigm shift from a mono- to pan- facility culture, it is worth noting that the majority of all photons produced, since mass-energy decoupling (by energy or number), arise in the ultra-violet, optical and near-IR regimes \citep[see the recent summary of the extra-galactic background by][]{Hill18}. Half of these photons are predominantly produced by stars and through star-formation, and the other half are produced through the accretion of baryonic material onto supermassive black holes. One important caveat, is that almost half of these freshly minted photons \citep{Dunne03,Driver16,Driver16b}, are almost immediately attenuated by dust grains, which reradiate the energy into the far-IR, before it emerges from the host galaxy. Including the shifting of wavelengths longward due to the expansion, the implication is that when building our panchromatic perspective, one might wish to start where photon production is dominant and readily detectable (i.e., the optical/near-IR) and where, arguably, the information content is highest. 

Here we describe the construction of a new deep optical/near-IR imaging dataset, built upon two ESO Public Surveys \citep[VST KiDS\footnote{Kilo Degree Survey} and VISTA VIKING][]{deJong13, Arnaboldi07} combined with the Galaxy And Mass Assembly (GAMA) panchromatic and spectroscopic survey \citep{Driver11, Hopkins13, Liske15, Driver16b}. 
In particular a key 50 sq degree region \citep[G23/WD23; see][]{Driver19}, will be targeted for future high-density spectroscopic, x-ray spectral, and radio line and radio continuum observations. 
The dataset presented herein, and including all unique GAMA redshifts, therefore forms the basis upon which to grow our panchromatic perspective. 

At its core the Galaxy And Mass Assembly Survey \citep[GAMA;][]{Driver11, Liske15}, spanning five fields, is a spectroscopic Legacy campaign using the Anglo Australian
Telescope's AAOmega wide-field facility \citep[see][]{Hopkins13}. The five fields are each $\sim$50--60 sq degrees in extent, and located at: 2$^h$ (G02), 9$^h$ (G09), 12$^h$ (G12), 14.5$^h$ (G15), and 23$^h$ (G23). The G09, G12 and G15 fields lie in the equatorial North Galactic Cap region, and have similar properties in terms of depth of the spectroscopic follow-up ($r<19.8$ mag), area (60 sq deg), spectroscopic completeness (98 per cent), and panchromatic coverage \citep[UV to far-IR;]{Driver16b}. 
The bulk of the GAMA science to date is based on the analysis of these three fields.
The original GAMA G02 field overlaps with the VIPERs and XMM-XXL equatorial field, and covers 55.7 sq degrees in extent. 
The field was not completed, however a 19.5 sq degree sub-region attained uniform 95.5 per cent spectroscopic completeness to $r<19.8$ mag \citep[see][]{Baldry18}. 

The final GAMA field at 23$^h$ and $-32.5\deg$, lies in the Southern Galactic cap, with a spectroscopic survey limit of, $i < 19.0$ mag, but with a slightly lower completeness of 94 per cent \citep[see][]{Liske15}. 
To date little science has been based on the G23 region \citep[although see studies such as][for examples where these data have been used]{Bilicki18, Vakili19}, however in due course it represents our premier field, because of its suitability for southern hemisphere follow up. In particular, this follow-up will be conducted by radio facilities, and a deep spectroscopic extension is planned as part of the Wide Area VISTA Extra-galactic Survey (WAVES); one of ten core surveys to be conducted by the 4MOST Consortium \citep[see][]{Driver19,deJong19}. 
This will extend the G23 region at high spectroscopic completeness ($>90$ per cent), to a limit of $m_Z \leq 21.2$. In addition WAVES will also survey the full KiDS region ($m_Z \leq 21.2$, $z_{\rm phot} < 0.2$), and the LSST Deep-Drill fields ($m_Z \leq 21.2$, $z_{\rm phot} < 0.8$).
Note that object selection for the WAVES-wide survey will be conducted using joint KiDS and VIKING photometry. 

In preparation, the G23/WD23 region is being extensively observed by Southern Hemisphere located radio facilities including: the Australian Compact Array (ATCA) as part of the GAMA Legacy ATCA Sky Survey (GLASS; Hyunh et al.~in prep.), the Australian Square Kilometre Array Pathfinder \citep{Leahy19} as part of the EMU\footnote{Evolutionary Map of the Universe} \citep{Norris11}, DINGO\footnote{Deep Investigation of Neutral Gas Origins} (Meyer et al.~in prep.) and FLASH\footnote{First Large Absorption Survey in H\,\textsc{i}} \citep{Allison20} surveys, and by the Murchison Wide-Field Array GOLD and MIDAS surveys (Seymour et al.~in prep). 
The expectation is that the G23/WD23 region, with its exceptionally high-density and deep spectroscopic completeness, should be a suitable location for a medium-deep survey, with upcoming facilities such as the Vera Rubin Observatory\footnote{Formerly referred to as the Large Survey Synoptic Telescope (LSST)}, the Square Kilometer Array (SKA), Euclid, and the Wide-Field InfraRed Space Telescope (WFIRST).

In terms of panchromatic imaging, G23/WD23 currently has comparable coverage to the GAMA equatorial fields, with data arising from concerted GALEX (NUV), VST KiDS, VISTA VIKING, WISE, and Herschel imaging campaigns. 
This wealth of data, combined with radio observations, and future upcoming deep spectroscopic observations, makes G23 a field of interest in coming years for extensive follow-up of either the entire field, or well selected sub-samples.

In Section \ref{sec:kidsdata} we describe the assimilation of the KiDS data into the GAMA Panchromatic Database followed by the generation of the base source catalogues from FUV to W2 using the new ProFound image analysis package \citep{Robotham18}. 
In Section \ref{sec:FIR} we use \textsc{ProFound} PSF-convolution mode to obtain photometry from W3 through to the PACS and SPIRE far-IR bands for objects brighter than $r \sim 20.5$ mag. In Section \ref{sec:verification} we verify the zeropoints, astrometry and compare our revised photometry to our previous LAMBDAR-based photometry.
This includes verification of the zeropoints and astrometry, star masking using GAIA DR2, Galactic extinction corrections using \textit{Planck}, star-galaxy separation based on colour and size, extensive visual inspection, and comparisons to earlier data. 
In Section \ref{sec:catAccess} we provide information on how to access the catalogues, and in particular provide some example extractions. Two companion papers describe the search for low surface brightness galaxies within the dataset (Turner et al. in prep), and the use of the panchromatic data to reconstruct the star-formation history of individual galaxies and sub-populations via a ``forensic"-style analysis (Bellstedt et al.~in prep). Further papers incorporating radio observations are in preparation.

All magnitudes reported here are in the AB system and when necessary we assume a cosmology with $H_0 = 70\,\rm{km}\,\rm{s}^{-1}\,\rm{Mpc}^{-1}$, $\Omega_m = 0.3$ and $\Omega_{\Lambda} = 0.7$.

\section{Assimilation of VST KiDS into the GAMA Panchromatic Database}
\label{sec:kidsdata}

The target catalogues to the GAMA spectroscopic campaign were built upon three distinct optical surveys: the Sloan Digital Sky Survey (SDSS; G09, G12 and G15), the Canada-France Legacy Survey (CFHTLS; G02), and the European Southern Observatory's VLT Survey Telescope's Kilo-degree Suvey (KiDS; G23).
For the equatorial regions the GAMA input catalogue is described in detail in \cite{Baldry10} and for the G02 region is described in \cite{Baldry18}.
The G23 input catalogue, used for the GAMA spectroscopic survey, has not been described previously and in brief
was constructed in 2014 based on initial pre-release VST KiDS data. These data have since undergone a number of revisions in terms of re-determination of the photometric zero-points, replacement of low quality data frames, and the filling in of data gaps as the KiDS team have honed their reduction and analysis pipelines. 
Nevertheless, our early KiDS analysis resulted in an $i$-band limited target catalogue ($i < 19.0$ mag), with star-galaxy separation based on table-matched near-infrared colours and size estimates, augmented with extensive and fairly ad hoc eyeball checks (based on selections designed to identify artefacts and ensure no galaxies were misclassified as stars). This initial input catalogue is available from the GAMA database and, while not ideal nor optimal, formed the basis for spectroscopic observations with the AAOmega
facility on the AAT from 2014 -- 2016 see \citet{Liske15}. 

Since this time the VST KiDS team has completed $ugri$ coverage of the three GAMA equatorial fields, and the G23/WD23 field. These data have recently been released as part of VST KiDS DR4\footnote{\url{http://kids.strw.leidenuniv.nl/}}, \citep{Kuijken19} and provide near complete coverage in all bands across the four primary GAMA fields. In one region of G23 the DR4 data is missing, however data exists from the earlier DR3 release and so we include these three fields. 
We are hence now in a position to redefine the GAMA base optical/near-IR catalogues in a uniform manner across our four primary regions. In doing so we create both deeper, and higher resolution imaging, from which we can derive more robust flux, size measurements and derived parameters (e.g., stellar masses, star-formation rates, and photometric redshifts). The main purpose of this paper is to provide a record of this replacement process --- a process akin to swapping the tablecloth on a fully laid table.

KiDS DR4 data are downloadable from the ESO archive, and come pre-SWARPed \citep{Bertin10} into 1 sq degree tiles. These are astrometrically and photometrically calibrated by the KiDS team using, initially, the Sloan Digitial Sky Survey in the North and 2MASS in the South, and with further supplementary calibration to GAIA DR2 $g$, as part of the final DR4 calibration process. Note that the DR4 data tiles as released, contain both a zero-point for each tile ({\sc PHOTZ}), reflecting the initial calibration, \textit{and}, a further zero-point offset ({\sc DMAG}) to adjust any derived flux measurements to the GAIA DR2 $g$-band system. Our initial action is therefore to modify all DR4tiles to absorb the {\sc DMAG} correction into the specified zero-points, by scaling the data. This is done to legitimately mosaic tiles, allowing for the construction of KiDS maps at any location and any size within the KiDS footprint.

We now follow the procedure outlined in the GAMA Panchromatic Data Release \citep[PDR;][]{Driver16b} and build large single SWARP \citep{Bertin10} images for each GAMA region. These (very) large mosaics are available via the Panchromatic SWARP Imager\footnote{\url{https://datacentral.org.au/services/cutout/}}
\citep{Driver16b}. We discuss the revised panchromatic depth of this imaging in Section \ref{sec:verification}. 
In total we SWARP 280 sq degree {\sc tile}s from KiDS, and also take the opportunity to rebuild our VIKING SWARPs using additional data amounting to 129,869 VISTA detectors (see Koushan et al. in prep. for details on the VISTA VIKING data). This comprises a total
data volume of 3.44TB and all mosaics are available via the URL indicated above and via the Public Data Central portal\footnote{\url{http://datacentral.org.au/}}. 
Images showing a visual comparison of KiDS and SDSS data are provided in figure 1 of Turner et al. (in prep).

\subsection{The adoption of ProFound for source detection --- a brief digression}

In constructing the PDR \citep{Driver16}, we made use of the original source detection as provided by the Sloan Digital Sky Survey Data Release 6, within our survey footprint. 
Following star-galaxy separation based on colour and size criteria \citep{Baldry10}, these data were used to define the GAMA input catalogue for the GAMA spectroscopic survey of the three equatorial fields, undertaken on the Anglo-Australian Telescope from 2011-2016 \citep[see][]{Liske15}. 
The original SDSS-derived equatorial input catalogue, was later table matched to our independent $r$-band catalogues, determined using {\sc Source Extractor} \citep{Bertin96}. 
{\sc Source Extractor} was then applied in dual band mode to determine forced aperture photometry from $u$ to $K_s$, using the elliptical apertures defined by {\sc Source Extractor}. 
Further table matching to independent catalogues in GALEX, WISE, and Herschel bands resulted in the far-UV to far-IR publicly available PDR dataset\footnote{\url{https://datacentral.org.au/services/cutout/}}. 
This catalogue was later superseded by flux measurements also based on our $r$-band defined apertures but now using the LAMBDAR in-house software \citep{Wright16} to measure forced photometric fluxes in all 21 bands (covering UV to IR
wavelengths) following convolution of the initial aperture with the
relevant facility point-spread function.

Throughout this process, a number of important lessons related to galaxy photometry emerged. Firstly, the undesirable reliance on table-matching to connect the SDSS input catalogue to our {\sc LAMBDAR} photometry, which inherently introduces errors due to different deblending outcomes between the {\sc SDSS Imaging Pipeline} and {\sc Source Extractor} methodologies. 
Secondly, issues arose around the integrity of the {\sc Source Extractor} aperture definitions. 
In particular {\sc Source Extractor}, like most detection algorithms, can be prone to bright galaxy fragmentation, and in some cases highly erroneous apertures, often due to the defined aperture following an isophotal bridge and looping round a nearby bright star --- these issues arise because of the difficulty in simultaneously measuring fluxes for both very bright and very faint sources.
Similar issues were also identified in our re-analysis of panchromatic photometry in the G10/COSMOS field \citep{Andrews17}. Following visual assessment of all GAMA apertures, via a citizen science project, it became apparent that typically 10 per cent of all apertures did not define the object to the desired level of accuracy.

This led to the development of a new source finding code,\footnote{Available on Github: \url{https://github.com/asgr/ProFound}} \textsc{ProFound} \citep{Robotham18}, with three important philosophical changes in source finding. 
Firstly, instead of using circular ({\sc SDSS Imaging Pipeline}), or elliptical ({\sc Source Extractor}) apertures, {\sc ProFound} acknowledges that most galaxies have irregular shapes, particularly as one probes to higher redshifts, fainter isophotal levels, and closer to the confusion limit. 
The {\sc ProFound} software identifies and preserves the initial isophote (or segment), which may be regular or irregular in shape. 
Secondly, we introduced the concept of segment-dilation (akin to a curve-of-growth) to obtain pseudo-total fluxes through the sequential addition of layers of pixels surrounding each segment. 
This process continues until the flux converges (in our case defined by a less than 5 per cent increase in flux), or the maximum number of allowable dilations is reached, resulting in pseudo-total magnitude estimates. 
Thirdly, {\sc Source Extractor} uses a hierarchical or nested-deblend process, by which derived elliptical apertures may overlap, and hence where flux can be double counted if not managed appropriately in later analysis. 
In the {\sc ProFound} software package, a watershed deblending approach is taken, where during the dilation process segments are not allowed to overlap --- i.e., all the flux in any one pixel is allocated to one object
only. 

\begin{figure}
\includegraphics[width=0.45\textwidth]{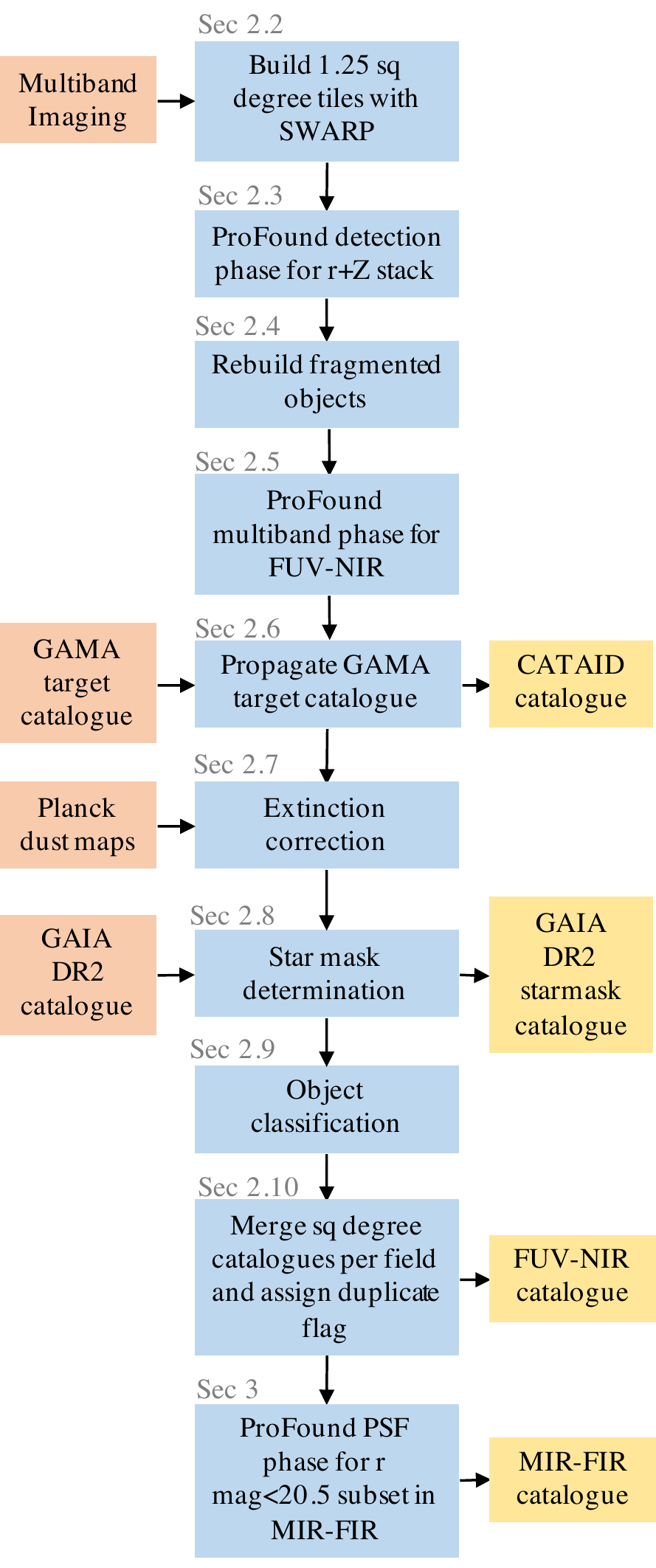}
\caption{Workflow for the adopted pipeline. Inputs to the pipeline are shown in orange, steps of the pipeline in blue, and outputs in yellow. }
\label{fig:flowchart}
\end{figure}

One can argue in specific cases as to which deblend approach, nested or watershed, is more appropriate, e.g., a nested approach is better for a small satellite within a large halo, while a watershed approach is better for dense complexes, or as one approaches the confusion limit. Our experience is that the watershed approach behaves better when things go pathologically wrong --- i.e., it is the least worst of the two approaches for difficult cases. 
For full details on {\sc ProFound} see the code description paper \citet{Robotham18}, or a recent applications to the deep DEVILS imaging data from VISTA VIDEO \citep{Davies18}. 
In the sections that follow we will adopt and apply ProFound and develop a pipeline around it to manage the multitude of issues that arise with wide area data collected from multiple ground-based facilities. We show the adopted workflow for the pipeline presented within this paper in Fig. \ref{fig:flowchart}.

\begin{figure*}
\includegraphics[width=\textwidth]{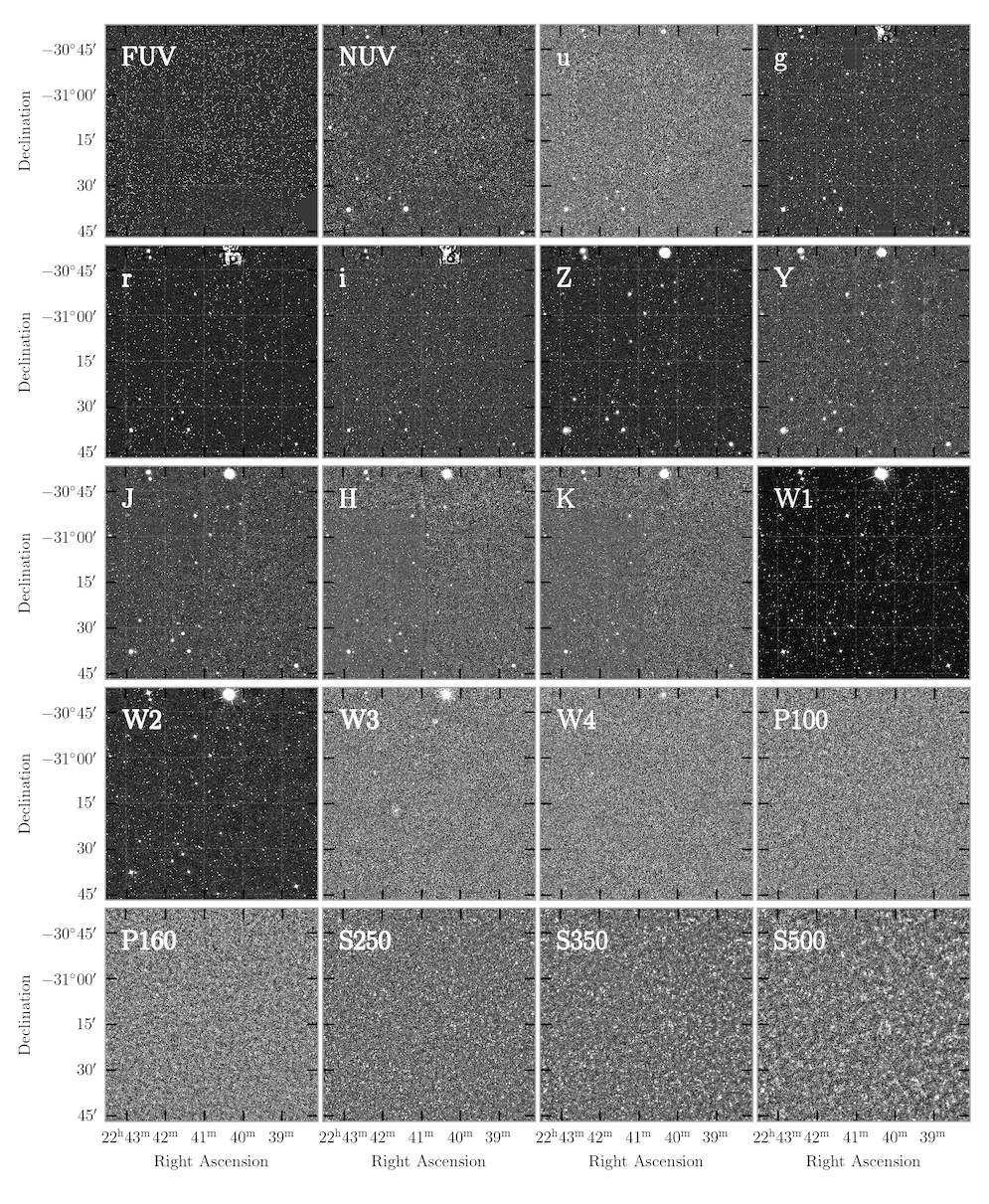}
\caption{Each frame shows a SWARPed $1.12^{\deg} \times 1.12^{\deg}$ {\sc tile} for the band indicated in the top left. Similar images were used to inspect the full dataset and record any pertinent issues, such as the missing detector in the $g$ {\sc TILE} above.}
\label{fig:swarp}
\end{figure*}

\begin{table*} 
	\begin{threeparttable}
		\centering
		\caption{Summary of key statistics for the imaging datasets used in each of the photometric bands used in this release. Limits provided by \citet{Driver16} indicate the range of values over the four GAMA fields.  \label{tab:SurveySummary}}
		\begin{tabular}{c|c|c|c|c|c|c|c} \hline \hline
			Band & Central Wavelength & Instrument & Data set/survey &Area & Limits  &  median seeing & Zero-point\\
			&  & & &(sq. deg) & (AB mag) &  (arcseconds) & (AB mag for 1 ADU)\\
			\hline
			$FUV$ & 1 539 \AA & GALEX &MIS+GO $^{\star}$ &186.74   & 24.59 -- 26.40 $^{\triangleright}$&  N/A & 18.82 \\
			$NUV$ & 2 316 \AA &  GALEX & MIS+GO $^{\star}$ &204.3  & 23.64 -- 24.07  $^{\triangleright}$&  N/A&  20.08\\
			$u$ & 3 582 \AA &  VST & KiDS $^{\ast}$ &211.21 & 24.8  $^{\ast}$ &  $0.9-1.1$ $^{\ast}$ & 0 \\
			$g$ & 4 760 \AA &  VST &KiDS $^{\ast}$ &211.21 & 25.4  $^{\ast}$ & $0.7-0.9$ $^{\ast}$ & 0 \\
			$r$ & 6 326 \AA &  VST & KiDS $^{\ast}$&211.21 & 25.2  $^{\ast}$ &$<0.6$ $^{\ast}$&  0\\
			$i$ & 7 599 \AA &  VST & KiDS $^{\ast}$&211.21 & 24.2  $^{\ast}$ &   $<1.1$ $^{\ast}$& 0 \\
			$Z$&8 854 \AA &  VISTA & VIKING $^{\dagger}$ &211.21 &  23.04 -- 23.19 $^{\triangleright}$ &  1.0 $^{\ddagger}$& 30 \\
			$Y$ &10 229 \AA &  VISTA & VIKING $^{\dagger}$ &211.21 &22.34 -- 22.51 $^{\triangleright}$ & 1.0 $^{\ddagger}$ & 30 \\
			$J$ & 12 556 \AA &  VISTA & VIKING $^{\dagger}$ &211.21 & 22.06 -- 22.21 $^{\triangleright}$ &  0.9 $^{\ddagger}$& 30 \\
			$H$ &16 499 \AA &  VISTA & VIKING $^{\dagger}$ &211.21 &  21.33 -- 21.42 $^{\triangleright}$ & 1.0 $^{\ddagger}$&  30\\
			$K_S$ & 21571 \AA &  VISTA & VIKING $^{\dagger}$  &211.21 &  21.30 -- 21.48 $^{\triangleright}$ &  0.9 $^{\ddagger}$&  30 \\
			$W1$ & 3.40 $\mu$m &  WISE & AllSky $^{\diamond}$ &211.21 & 21.09 -- 21.41 $^{\triangleright}$ &  N/A& 23.16 \\
			$W2$ & 4.65 $\mu$m  &  WISE & AllSky $^{\diamond}$ &211.21 &  20.26 -- 20.77 $^{\triangleright}$ & N/A & 22.82 \\
			$W3$ & 12.8 $\mu$m  &  WISE & AllSky $^{\diamond}$&211.21 & 18.44 -- 18.89 $^{\triangleright}$ &  N/A& 23.24 \\
			$W4$ & 22.4 $\mu$m &  WISE & AllSky $^{\diamond}$ &211.21 & 16.54 -- 16.96 $^{\triangleright}$&  N/A& 19.6 \\
			$P100$  & 98.9 $\mu$m  & PACS & ATLAS $^{\triangleleft}$&211.21  &  12.96 -- 13.14 $^{\triangleright}$ &  N/A&  8.9\\
			$P160$ &  156 $\mu$m &  PACS &  ATLAS $^{\triangleleft}$&211.21& 13.44 -- 13.66 $^{\triangleright}$&   N/A& 8.9 \\
			$S250$ & 249 $\mu$m  &  SPIRE & ATLAS $^{\triangleleft}$&181.14 & 12.52 -- 12.60 $^{\triangleright}$ &  N/A& 11.68 \\
			$S350$ & 350 $\mu$m  &  SPIRE & ATLAS $^{\triangleleft}$&181.14 & 12.36 -- 12.51 $^{\triangleright}$ & N/A&  11.67\\
			$S500$ & 504 $\mu$m  &  SPIRE & ATLAS $^{\triangleleft}$&181.14 &  12.16 -- 12.23 $^{\triangleright}$ &  N/A&  11.62\\
			\hline
		\end{tabular}
		\begin{tablenotes}
			\small
			\item $^{\star}$ \citet{Martin05}, $^{\ast}$ \citet{deJong13, deJong13b}, $^{\dagger}$ \citet{Edge13}, $^{\ddagger}$ \citet{Venemans15}, $^{\diamond}$ \citet{Wright10}, $^{\triangleleft}$ \citet{Eales10}, $^{\triangleright}$ \citet{Driver16}
		\end{tablenotes}
	\end{threeparttable}
\end{table*}

\subsection{Building 1.25 sq degree overlapping tiles with {\sc SWARP}}

Prior to running {\sc ProFound} on KiDS and VIKING data, we first use the {\sc SWARP} package \citep{Bertin10} to build slightly extended $1.12^{\deg} \times 1.12^{\deg}$ images in each band ($FUV$, $NUV$, $u$, $g$, $r$, $i$, $Z$, $Y$, $J$, $H$, $K_s$, $W1$, $W2$, $W3$, $W4$, $P100$, $P160$, $S250$, $S350$, $S500$) - essentially adding overlap regions. 
These revised tiles are centred on the rescaled KiDS 1 sq degree tiles sourced from the ESO archive. 
In {\sc SWARP}ing the data we regrid all bands to a resolution of 0.339 arcsec, use a background smoothing mesh of $256 \times 256$ (pixels), and a background filter-size of $3 \times 3$ (background cells). The dat frames are combined using the MEDIAN combine option, which we have deemed to be the most stable option for regions with poor-quality data. 

Fig.~\ref{fig:swarp} shows an example of a SWARped tile indicating the depth and quality in each band. Visual inspection was made of all data frames, using images similar to Fig.~\ref{fig:swarp}, to ensure each tile in each band was correctly built. 
In some cases it was noted that the individual tiles provided by KiDS were missing detectors in some bands, and in some tiles, which will leave gaps in the panchromatic coverage. Objects with missing coverage in a particular band will have fluxes and magnitudes set to either NA or -999. 
Based on a visual inspection of the image quality, we conclude that a {\sc ProFound} analysis based on combined $r+Z$ stacks will provide near complete and contiguous coverage over all four GAMA regions.

\begin{figure}
\includegraphics[width=0.47\textwidth]{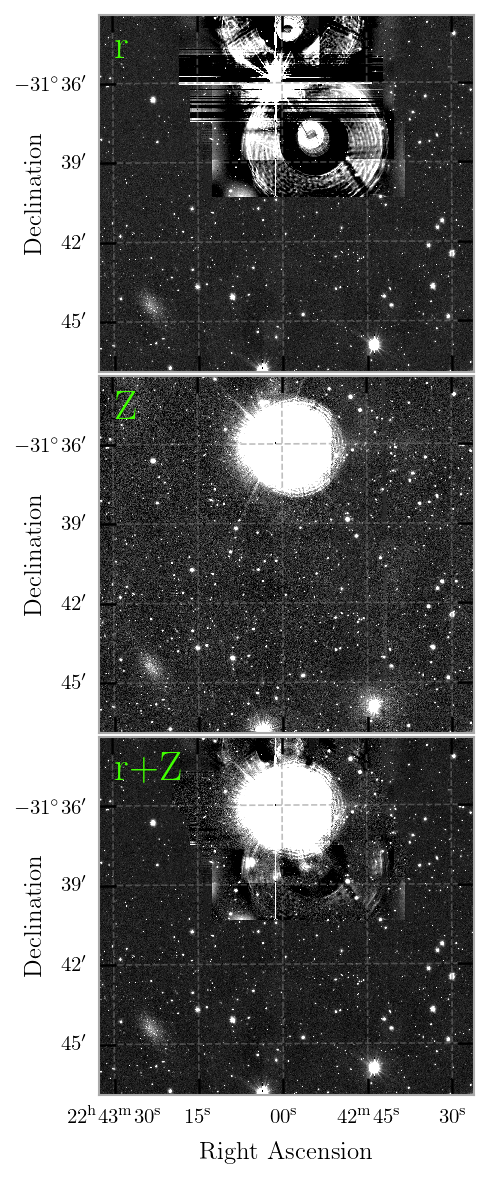}
\caption{The above panel shows a section from the top mid region of Fig.~\ref{fig:swarp} VST KiDS $r$ frame (top) and a VISTA
  VIKING $Z$ frame (middle) and the combined stack weighted by inverse
  noise variance (bottom).  A problem region is shown indicating how
  the use of two frames from different telescopes helps to mitigate
  issues due to stellar ghosting.}
\label{fig:zrzr}
\end{figure}

\subsection{Source detection with {\sc ProFound}}
\label{sec:sourceDetection}

Source detection is conducted via the {\sc ProFound} package \citep{Robotham18}. This is based on \textsc{ProFound} version 1.10.8 which can be obtained from \url{https://github.com/asgr/ProFound}. 
Within {\sc ProFound} numerous parameters exist that determine the manner in which sources are extracted from the image. 
For completeness we show the exact command we use in the Appendix using the {\sc profoundMultiBand} command. 
The command allows the user to specify one or more bands to use for the detection pass as well as the bands for which measurements should be made. In the initial source detection phase, only the detection bands are provided. 
When multiple images are specified for detection they are combined in an inverse variance weighting based on the internal background assessment. Here we combine data from the KiDS r band and the VIKING Z band, i.e., $r+Z$. This has the distinct advantage of overcoming some artefact effects such as ghosting, satellite trails, and bad pixels. 
Fig.~\ref{fig:zrzr} shows a KiDS VST $r$ band image (upper), a VISTA VIKING image (middle), and the combined $rZ$ image (lower).

\begin{figure*}
\includegraphics[width=\textwidth]{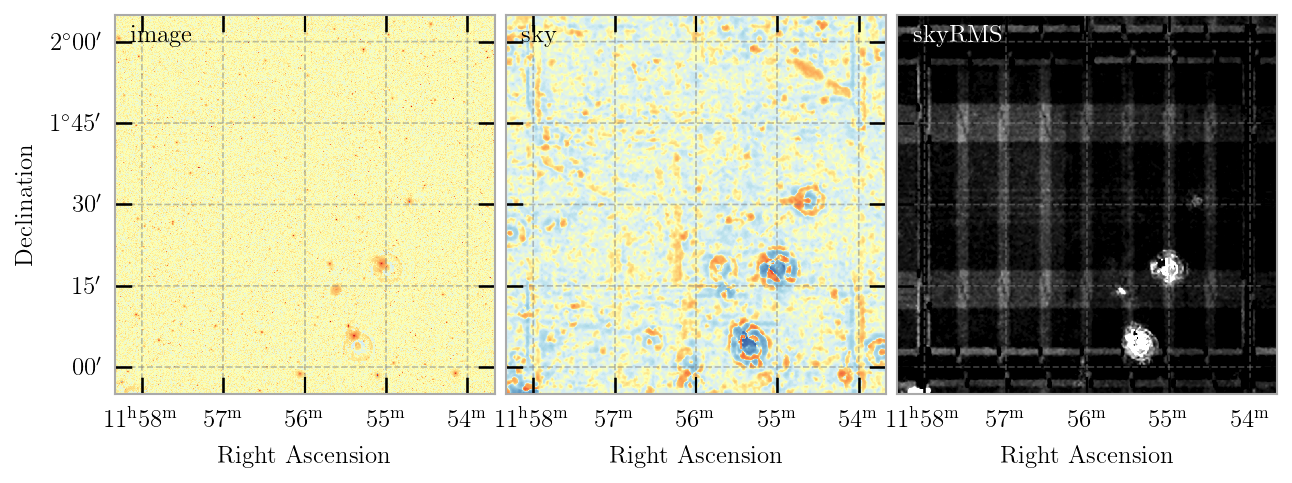}
\caption{ProFound's inbuilt background analysis, showing the original sq degree data (left), the derived background sky map (centre), and the derived background root mean square map (right).}
\label{fig:back}
\end{figure*}

Inherent to {\sc ProFound}, is its robust modelling of the local background, which may include the sky, the haloes of bright objects, scattered light or artificially enhanced regions, through appropriate median filtering \citep[see][]{Robotham18}. 
Fig. ~\ref{fig:back} shows an example {\sc tile} with the original image (left), the derived background map (centre, where the background is seen to be elevated near the positions of bright objects, particularly bright ghosting), and the sky
root-mean sky statistics (right, again indicating regions of heightened uncertainty in the background estimate). 
This information is used in determining flux errors, ensuring objects in noisier regions have appropriately derived errors. 
In examining the background in detail the genesis of the data is also apparent in the Sky Root Mean Square (SkyRMS) map (right panel). 
This is a common feature in surveys we have studied, and in this case highlights the varying noise characteristics of the individual detectors going into the initial {\sc SWARP} {\sc tile} image.

\subsection{Rebuilding fragmented galaxies}
\label{sec:rebuildingFragmentedGalaxies}

A common problem in most automated detection algorithms is that of fragmenting of bright galaxies. To check this we select all known galaxies from the Third Reference Catalogue \citep[RC3;][]{deVauc91} which lie within the GAMA regions, and produce cutout images with the derived segments overlain. There are 21, 144, 31, and 10 RC3 galaxies within the G09, G12, G15, and G23 regions respectively. 
Note that the G12 region includes the Virgo Southern Spur, and G23 includes a nearby void region. Initial investigations showed that {\sc ProFound} also tended to overly fragment very bright galaxies. 
Two enhancements were implement in {\sc ProFound} to assist with this.

One new parameter on top of the standard \textsc{tolerance} threshold (which determines how much peak flux an object needs relative to neighbouring objects before being merged) is \textsc{reltol}. 
This modifies the \textsc{tolerance} by the ratio between the segment peak flux and the saddle point flux where it touches a neighbouring segment to the power of \textsc{reltol}. Since the default is \textsc{reltol}$=0$, this will in general have no effect. 
However, when it is made larger than 0 merging becomes more aggressive in the outskirts of galaxies where the peak flux will tend to be much larger than the saddle point flux.
Subjectively, raising this above 0 tends to do a better job of keeping very extended and flocculent spiral galaxies intact, and it has little negative impact on the fainter source deblending that parameters will tend to be optimised for (since this is where most of our survey sources exist).

The other new parameter to better control segmentation is \textsc{cliptol}. 
This specifies the saddle point flux above which segments are always merged, regardless of competing criteria. 
For very bright objects with complex image artefacts (e.g.\ around bright stars) this proves to be very successful at properly reconstructing sources which might otherwise be significantly fragmented due to the presence of spurious flux discontinuities. 
Given most applications will apply a bright star mask, this option is perhaps somewhat cosmetic, but it does mean \textsc{ProFound} will return reasonable photometry even for the brightest and most difficult sources.

\begin{figure*}
\includegraphics[width=\textwidth]{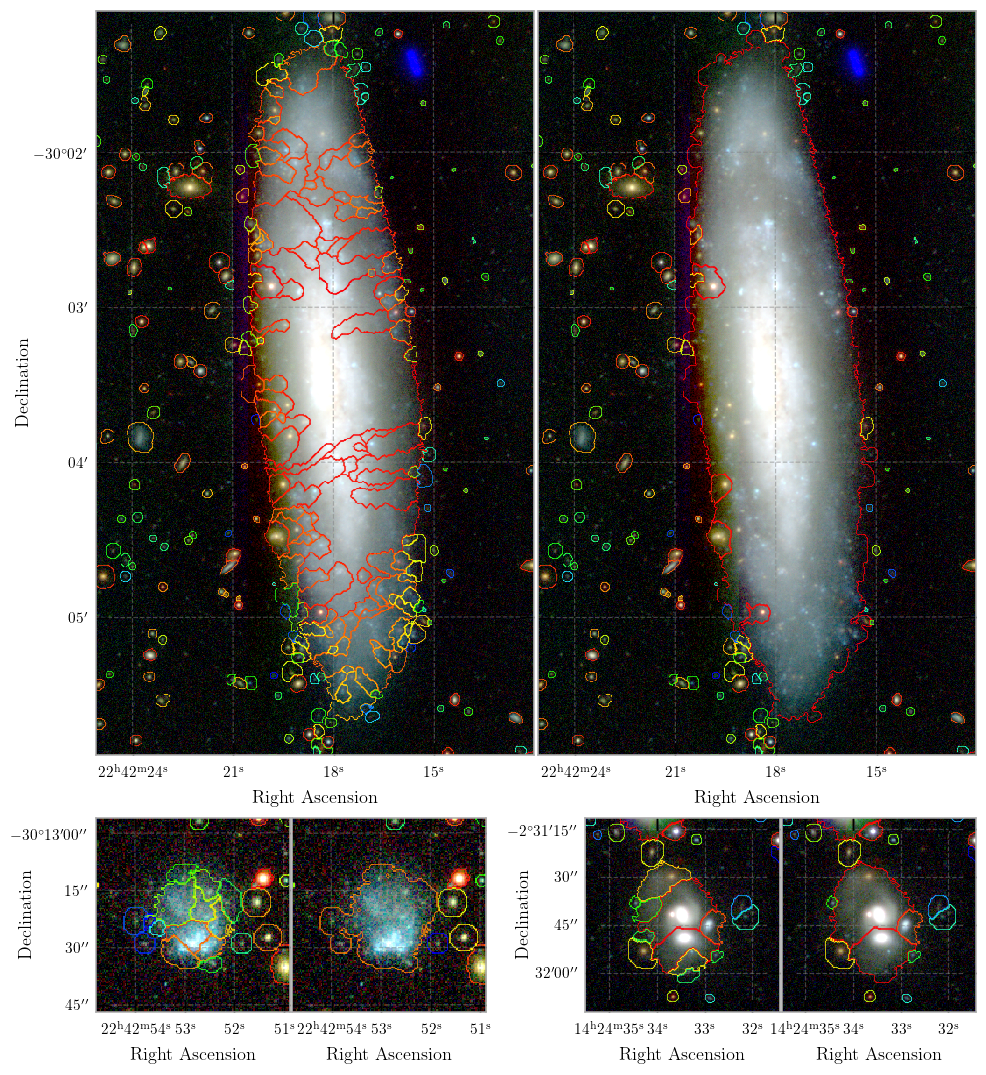}
\caption{Three examples of objects whose segments have been regrouped using the \texttt{profoundSegimFix} function within \textsc{ProFound}. Note how the severity of fragmentation in these examples varies. For each example, the left plot shows the initial segmentation, whereas the right plot shows the fixed segmentation. }
\label{fig:frag}
\end{figure*}

As this fragmentation was occurring despite the new \textsc{ProFound} parameters, a process was implemented to manually regroup segments that belong to a single object. 
To complete this task, an in-house tool was developed that allowed users to view a thumbnail of an object and click on segments to be regrouped. This tool is available through the \texttt{profoundSegimFix} function within \textsc{ProFound}. An example of an object whose segments have been merged in this way can be seen in Fig. \ref{fig:frag}, where the right panel shows the resulting segmentation map after merging. Per square degree, an output file was produced that recorded which segments (as determined by the detection phase of \textsc{ProFound}) needed to be regrouped. 

Rather than visually inspecting every object, only objects with three or more abutting segments, a total group\footnote{In the context of the photometry, \textit{group} refers to a collection of segments directly abutting each other. } magnitude $<20.5$, and groups not flagged as containing a star were selected to be visually checked. For the full GAMA sample, this resulted in 75,863 objects. Of these, 6,777 required manual intervention. In total, this task took a week, with seven authors (SB, SPD, AR, LJD, JT, RC, KB, HH) assisting in the regrouping process. 

Before running the multiband form of \textsc{ProFound}, the manual fixes to segments were applied to the detected segmentation map using the command \texttt{profoundSegimKeep}. This fixed segmentation map was used for the remainder of the photometry pipeline. 

\subsection{Multiband Photometry with {\sc ProFound}}
\label{sec:multibandPhotometry}

After initial segments (isophotal outlines), are defined from the stacked $r$ and $Z$ image (as described in Section \ref{sec:sourceDetection}) and have been fixed for fragmentation (as described in Section \ref{sec:rebuildingFragmentedGalaxies}), the fixed segments then form the basis for subsequent measurements in the analysis bands ($FUV$, $NUV$, $u$, $g$, $r$, $i$, $Z$, $Y$, $J$, $H$, $K_s$, $W1$, $W2$). 

Within each band, flux measurements are presented in two different ways. In order to account for all of the flux of a single object, individual segments require dilation beyond the detected segment. 
Dilation is conducted iteratively, where the edges are extended until all object flux has been accounted for. 
The background sky estimate is made as both a \textit{global} sky measurement, or a \textit{local} sky measurement, where the sky is measured within the dilated annulus. 
The resulting flux when using the global sky measurement for sky subtraction is indicated as \textsc{flux\_t} in the catalogue, whereas the flux resulting from a local sky subtraction is indicated as \textsc{flux\_l}. 
For large objects with significant halo flux, a local sky subtraction is liable to subtracting off the halo light, and therefore the \textsc{flux\_t} measurement is expected to better represent the total galaxy flux. 
Conversely, for very faint galaxies (particularly those that are close to noisy regions), we expect that the \textsc{flux\_l} measurement will best represent the galaxy flux. 
For the sake of selection cuts in the remainder of the paper, we have utilised the \textsc{flux\_t} measurement. 

The flux errors derived by \textsc{ProFound} include the errors due to sky subtraction and sky RMS (as outlined in Section \ref{sec:sourceDetection}). Whilst it is also possible for \textsc{ProFound} to derive the error contribution by pixel noise correlation (introduced through the SWARP pixel resampling process), we have not included this contribution in our analysis as it contributes only a small (often negligible) fraction of the error introduced by the sky and sky rms. Based on tests for a single square degree, we find that the median pixel correlation contribution to the flux error is highest in the $NUV$/$W1$/$W2$ bands with 11/12/8 per cent respectively, but overall it is much smaller, with a mean contribution of 3 per cent over all bands. The error due to pixel correlation is expensive to compute, therefore omitting this uncertainty contibution saves significant computational time. We refer the reader to \citet{Robotham18} for the details on uncertainty derivation in \textsc{ProFound}. 

For completeness, we show the \texttt{profoundMultiBand} command used to derive our multi-band catalogues for each SWARped tile in the Appendix. This amounts to the production of 280 distinct catalogues each containing around 300,000 objects and taking about 12hrs to build the SWARPs, and a further 6hrs to process \textsc{ProFound}. The software is hence run on the Pawsey Supercomputing Centre's Zeus machine, taking about 2 days to complete a full run across all tiles.

\subsection{Linking to GAMA objects}
\label{sec:GAMAmatch}

To link the new photometric catalogue to the existing GAMA target catalogue (which contains 1,468,620 objects across all four GAMA fields), we project the GAMA catalogue onto the \textsc{ProFound} segments. 
A successful projection will occur if a GAMA coordinate is encompassed by a corresponding \textsc{ProFound} segment. In some cases (44,766 instances over all four fields), a single \textsc{ProFound} segment is linked to more than one GAMA input object. This generally occurs if multiple GAMA targets were placed on a single object, but can also occur if two objects have not been appropriately deblended by \textsc{ProFound}, and hence share a segment. 
Two strategies are implemented in order to decide which GAMA ID should be assigned to the segment when a single segment coincides with multiple GAMA objects. If the GAMA objects have spectroscopically-measured redshifts, then the selected ID is taken to be the object whose redshift is closest to the flux-weighted mean redshift of all GAMA objects present in the segment. 
If redshift measurements do not exist, however, then the selected ID is taken from the object contributing the largest amount of flux to the segment. 

For individual cases where a segment contains multiple GAMA sources that have redshifts varying by more than 0.1, we allocate a flag \textsc{Z\_ConfusionFlag}$=1$. This assists in the identification of objects for which the redshift measurement is not indicative of all the flux in the segment. 

\subsubsection{Objects in KiDS/VIKING not in GAMA}
After matching our final KiDS/VIKING catalogue to the GAMA input catalogue (which extends to $r_{\rm GAMA}$=21.0 mag), we cut the catalogue at $r_{\rm KiDS} < 20$ mag (just beyond the GAMA spectroscopic limit of $m_r = 19.8$).
This identifies almost 11,000 galaxies not previously recorded. Visual inspection of all $\sim$11,000 reveals that $\sim$6,000 of these objects are artefacts not previously flagged, $\sim$3,000 are galaxies not previously identified, and the remaining objects are equally divided between stars or ambiguous objects. In total these objects represent $<1$ per cent of the galaxies above this flux limit but nevertheless we introduce an eyeball flag {\sc eyeclass} so that these objects can be indicated. We also introduce an {\sc uberclass} flag which takes as its value the {\sc eyeclass} if known or the {\sc class} flag if not known. We therefore recommend the {\sc uberclass} flag be used to extract star, galaxy and/or ambiguous subsets. 
In addition to the previously missed galaxies we identify a mixture of low surface brightness systems, and objects that have been identified as the close pair of a previously-identified galaxy, but had not been separately resolved in the past. 
In a companion paper, Turner et al. (in prep), we provide more detail on these objects and discuss the implications for the stellar mass density.

\subsubsection{Objects in GAMA not in KiDS/VIKING}
Similarly we can also identify objects in the GAMA spectroscopic target catalogue ($r_{\rm GAMA}<19.8$mag) that are not matched in the new KiDS/VIKING catalogues. Either, a non-match arises from the fact that no object has been identified at the coordinate of the GAMA object, or because multiple GAMA objects have been engulfed by a single segment, resulting in only a portion of the GAMA objects appearing in the final catalogues. 
Within the four GAMA fields, 16,068 objects have been identified for which no object has been detected in the KiDS/VIKING photometry, and in almost all cases this is because the GAMA object points to a sky position in our updated photometry. This is likely an indication that the original SDSS photometry on which the GAMA input catalogue was based contained some sort of artefact at these coordinates. Only $\sim$150 of these GAMA targets have securely measured redshifts, and these objects fall within regions that are missing imaging in $r+Z$, and hence do not appear in the new KiDS/VIKING catalogues. 
Finally, 44,766 objects have been identified within the new catalogues that match back to two or more GAMA targets, corresponding to $\sim3$ per cent of the total sample. As a result, an additional 44,782 GAMA targets do not appear in the KiDS/VIKING catalogues. 
For those cases where multiple GAMA targets in a single \textsc{ProFound} segment had redshifts, we find that in 69 per cent of cases have $\Delta z<0.01$, implying that in the majority of cases where GAMA targets have been consolidated, these do in fact belong to a single object along the line-of-sight. 
This highlights that the original GAMA target catalogue had fragmented objects more often than \textsc{ProFound} has merged multiple objects into a single segment. 
Hence, a total of $\sim4$ per cent of the GAMA target objects do not appear in our updated catalogues. 

\subsection{Extinction corrections using Planck E(B-V) maps}
\label{sec:extinctionCorrection}

\begin{figure*}
\includegraphics[width=0.95\textwidth]{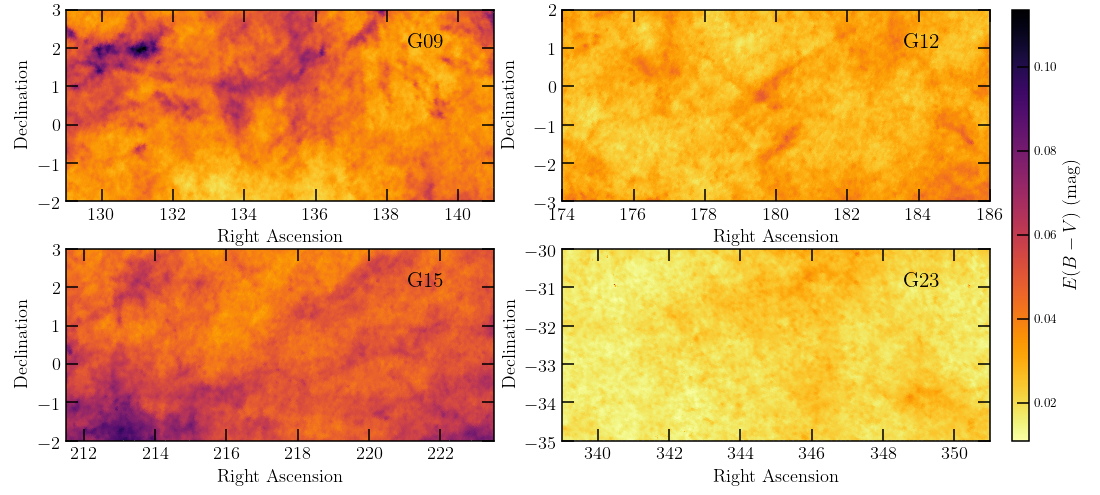}
\caption{Galactic Extinction, based on the Planck E(B-V) values, in
  the four GAMA fields, clockwise from top left are G09, G12, G23 and
  G15. The G23 field has the lowest extinction.}
\label{fig:extinction}
\end{figure*}

To correct our magnitudes for the effects of Galactic extinction we
use the Planck $E(B-V)$ map\footnote{\textsc{HFI\_CompMap\_ThermalDustModel\_2048\_R1.20.fits}, \url{https://irsa.ipac.caltech.edu/data/Planck/release_1/all-sky-maps/previews/HFI_CompMap_ThermalDustModel_2048_R1.20/index.html}} \citep{Planck13}.
From this map we extract the
$E(B-V)$ values, and convert the HEALpix values to RA and Dec, and
identify the closest $E(B-V)$ value to each object in each of our catalogues. We then correct all
magnitudes, magnitude errors, surface brightnesses, surface brightness errors, fluxes and flux-errors for {\it all}
objects (i.e., stars, galaxies, artefacts etc). We determine the attenuation correction for each band in the normal way ($A_x = [A_x/E(B-V)] \times E(B-V)$) using the extinction co-efficients listed and cited in Table.~\ref{tab:atten} (which implicitly use the Galactic extinction law from \citealt{Schlafly11}). 

\begin{table} 
\begin{center}
\caption{Attenuation values used here in conjunction with Planck $E(B-V)$ values to determine extinction corrections for each line of sight and in each filter. Median $A_{x}$ values are determined by convolving the nearby galaxy templates from \citet{Brown14} with the filter response curves, and assuming either a \citet{Cardelli89} ($FUV-K_s$) or \citet{Fitzpatrick99} (WISE) Galatic Extinction Law with $R_v=3.1$. \label{tab:atten}}
\begin{tabular}{c|c|c} \hline \hline
Filter (x) & [$A_x/E(B-V)$] & Vega to AB \\ \hline 
\multicolumn{3}{c}{GALEX} \\ \hline
$FUV$ & 8.24152 & - \\
$NUV$ & 8.20733 & - \\ \hline
\multicolumn{3}{c}{ESO VST Omegacam$^{\dagger}$} \\ \hline
$u$ & 4.81139 & - \\
$g$ & 3.66469 & - \\
$r$ & 2.65460 & - \\
$i$ & 2.07472 & - \\ \hline
\multicolumn{3}{c}{ESO VISTA VIRcam$^{\ddagger}$} \\ \hline
$Z$ & 1.55222 & 0.502\\
$Y$ & 1.21291 & 0.600 \\
$J$ & 0.87624 & 0.916 \\
$H$ & 0.56580 & 1.366 \\
$K_s$ &  0.36888 & 1.827\\ \hline
\multicolumn{3}{c}{WISE} \\ \hline
$W1$ & 0.20124 & - \\ 
$W2$ & 0.13977 & - \\
$W3$ & 0.05433 & - \\
$W4$ & 0.02720 & - \\ \hline
\end{tabular}
\end{center}

$^{\dagger}$ \citet{Kuijken19} 

$^{\ddagger}$ \citet{Gonzalez-Fernandez18}
\end{table}

Fig.~\ref{fig:extinction} shows how the $E(B-V)$ values vary across
the four GAMA fields highlighting the significant structure due to
streaks of Galactic cirrus. However we note the maximum $E(B-V)$ shown
in the plots reaches only to 0.07, hence amounting to $0.2$ mag of
extinction in the $r$ band.

\subsection{Constructing the star-mask from GAIA DR2}
\label{sec:starMask}

\begin{figure}
\includegraphics[width=\columnwidth]{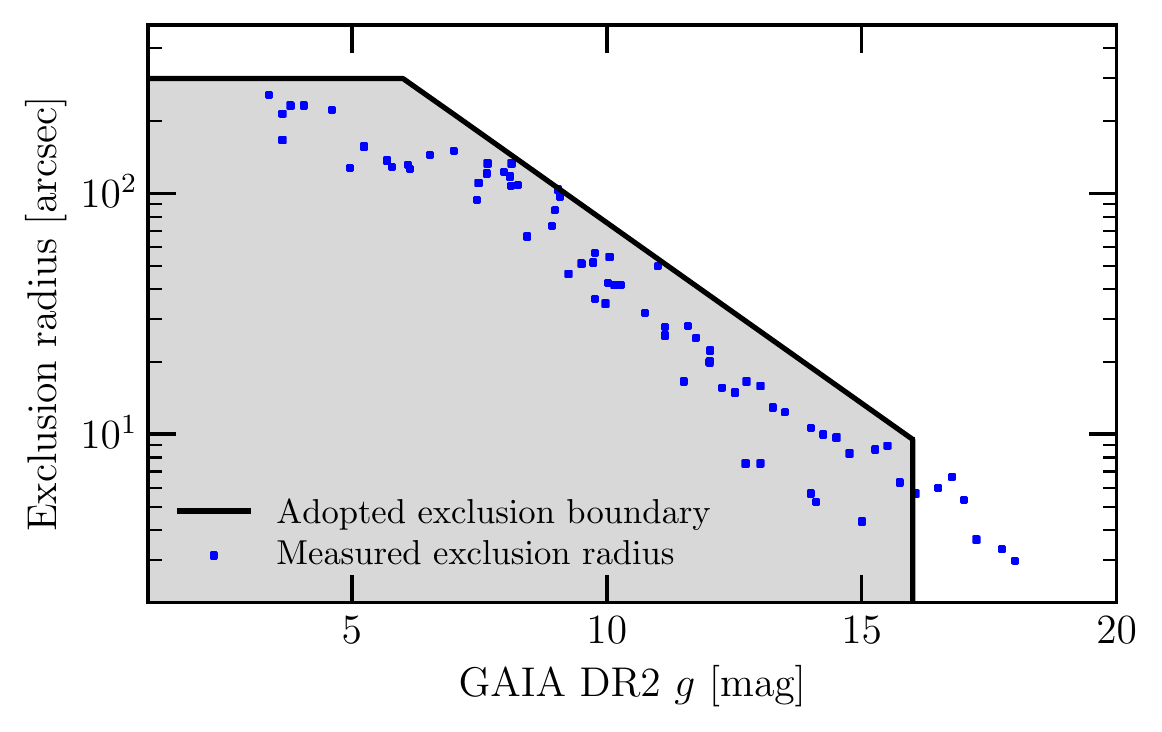}
\caption{The starmask exclusion radius (solid black line) was determined from eyeball measurements of randomly selected stars (blue data points) drawn from the four GAMA regions.}
\label{fig:starmask}
\end{figure}

Figs.~\ref{fig:swarp}, \ref{fig:zrzr} and ~\ref{fig:back} highlight the issue of ghosting around bright stars, and how the location of this ghosting is dependent on both the position within the focal plane, and the flux of these stars. 
Photometry of objects in these regions will be compromised, and for many purposes it will be necessary, or desirable, for these objects to be removed. 
To build a star-mask flag we elect to use the recently released GAIA DR2 catalogue \citep{Gaia18}, which contains robust positions for all bright objects across the sky. 
However, first we need to remove any galaxies in the GAIA DR2 catalogue, as we do not wish to mask these objects. 
To do this we match to both the RC3 catalogue, and also our previous GAMA catalogue, which has been extensively visually inspected and for which most objects (98 per cent), have had redshifts measured to $r<19.8$ mag (or redshifts measured out to $r<19.2$ mag in the case of G23). 
Matching GAIA DR2 to RC3 results in 54 matches within the GAMA regions, while matching to GAMA identifies a further 684 objects to GAIA DR2 $g < 18.0$ mag. 
These are removed from our GAIA star-mask catalogue. 
We then extract cutouts of a random sample of GAIA stars and identify a $g$ mag-radius relation, as shown in Fig.~\ref{fig:starmask}. 
Inside the radius indicated (solid black line), the artefact rate is extremely high, and photometry will be compromised and classification problematic.

\begin{figure}

\includegraphics[width=0.48\columnwidth]{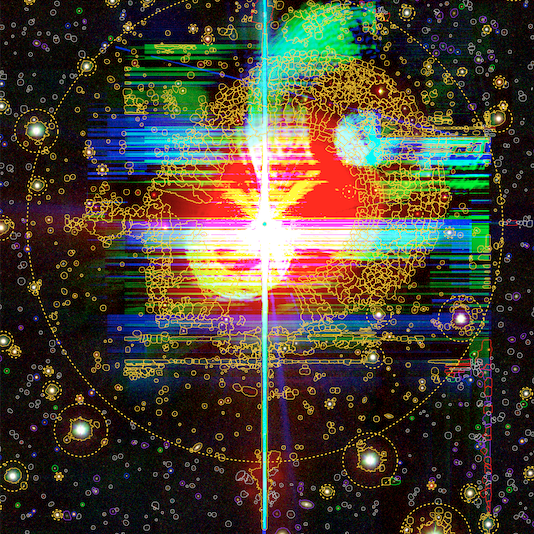}
\includegraphics[width=0.48\columnwidth]{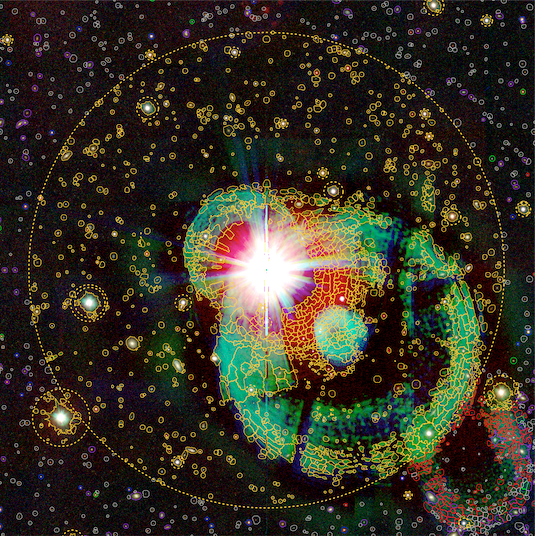}

\includegraphics[width=0.48\columnwidth]{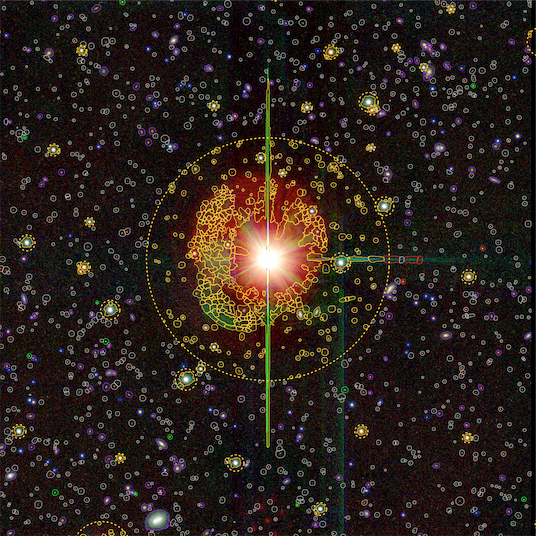}
\includegraphics[width=0.48\columnwidth]{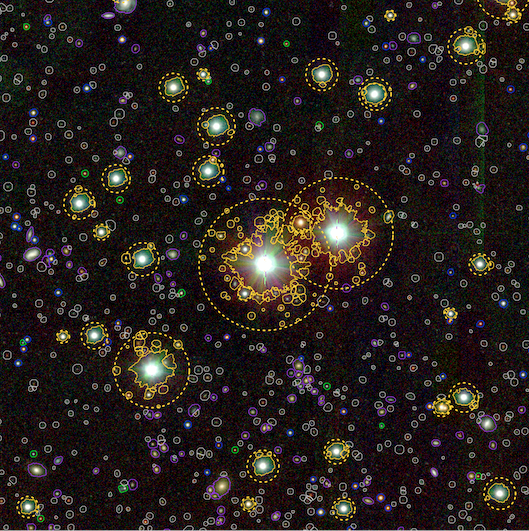}

\caption{Four panels showing our star mask regions, as indicated by the dotted lines. These lines are defined by Eqn.~\ref{eqn:starradius}. Objects within the star mask have their {\sc starmask} flag set to 1 and are shown with yellow contours. Upper are two of our brightest stars with $g_{\rm GAIA DR2}$ mag = 6.6 )left) and 8.0 mag (right). The lower panels show two more typical regions centred on stars with $g_{\rm GAIA DR2}$ = $9^{th}$ mag (left) and $11^{th}$ mag (right) but also showing stars extending to our cutout limit of $g_{\rm GAIA DR2}=16^{th}$ mag}
\label{fig:stars}
\end{figure}

Fig.~\ref{fig:stars} shows examples for four regions centred on four stars, showing firstly two extremely bright stars (which are relatively rare: upper), to two more typical regions with a smattering of masked stars (lower). The dotted circles indicate the masked region indicated by Eqn.~\ref{eqn:starradius}, and all objects within these regions have their {\sc starmask} flag set to 1 and are shown on these figures with yellow outlines.
\begin{eqnarray}
r[\mbox{arcmin}]=10^{(1.6-0.15g)}
\mbox{ and } [r < 5.0', g < 16.0]
\label{eqn:starradius}
\end{eqnarray}
Note that we only define a starmask around stars brighter than 16$^{th}$ mag, as below this the ghosting appears to lie below the sky noise. 
Fig.~\ref{fig:stars} highlights the exclusion zones around stars of various magnitudes (as indicated by the yellow segments). One can see that by 16$^{th}$ magnitude there is no need for exclusion regions.
Fig~\ref{fig:tiles} (centre panel) shows objects with {\sc starmask}=1 in blue, highlighting the foreground coverage lost due to bright stars. To determine the reduction in area we create a grid of equally spaced points at $6''$ intervals and apply our starmask criteria. 
We sum the grid-points within our GAMA boundary for which {\sc starmask}=0. 
This results in the areas as indicated in Table.~\ref{tab:areas}.

\begin{table*}
\caption{Locations of the four primary GAMA fields, their complete
  area coverage on the sky and the reduced are after subtracting of
  the area lost to the star-mask, and their astrometric offsets with
  respect to GAIA DR2.  \label{tab:areas}}
\begin{center}
\begin{tabular}{c|c|c|c|c|c|c|c} \hline \hline
GAMA & RA range & Dec range & Full area & Eff. area & Masked area & $\Delta RA$[GAMA-GAIA] & $\Delta Dec$[GAMA-GAIA] \\ 
field & (deg) & (deg) & (sq. deg) & (sq. deg) & (sq. deg) & (arcsec) & (arcsec) \\ \hline
G09 & 129.0 --- 141.0 & -2 --- +3   & 59.97 & 54.93 & 4.91 & 0.056 & 0.066 \\
G12 & 174.0 --- 186.0 & -3 --- +2   & 59.97 & 57.44 & 2.39 & 0.134 & 0.106 \\
G15 & 211.5 --- 223.5 & -2 --- +3   & 59.97 & 56.93 & 2.90 & 0.101 & 0.098 \\
G23 & 339.0 --- 351.0 & -35 --- -30 & 50.58 & 48.24 & 2.28 & -0.113 & 0.134 \\ \hline
\end{tabular}
\end{center}
\end{table*}

\subsection{Object classification}
\label{sec:objectClassification}

\subsubsection{Star, galaxy, and ambiguous classification}

Star-galaxy classification is performed in an initial phase using measured parameters, and then classifications are later overridden in a series of steps given prior knowledge (e.g., a known redshift, or direct eyeball classification). 

In the first round all objects are assigned a {\sc class} flag that is initially set to \texttt{ambiguous}, and therefore those objects not reclassified in the latter stages will retain an \texttt{ambiguous} flag.

\begin{figure}
\includegraphics[width=\columnwidth]{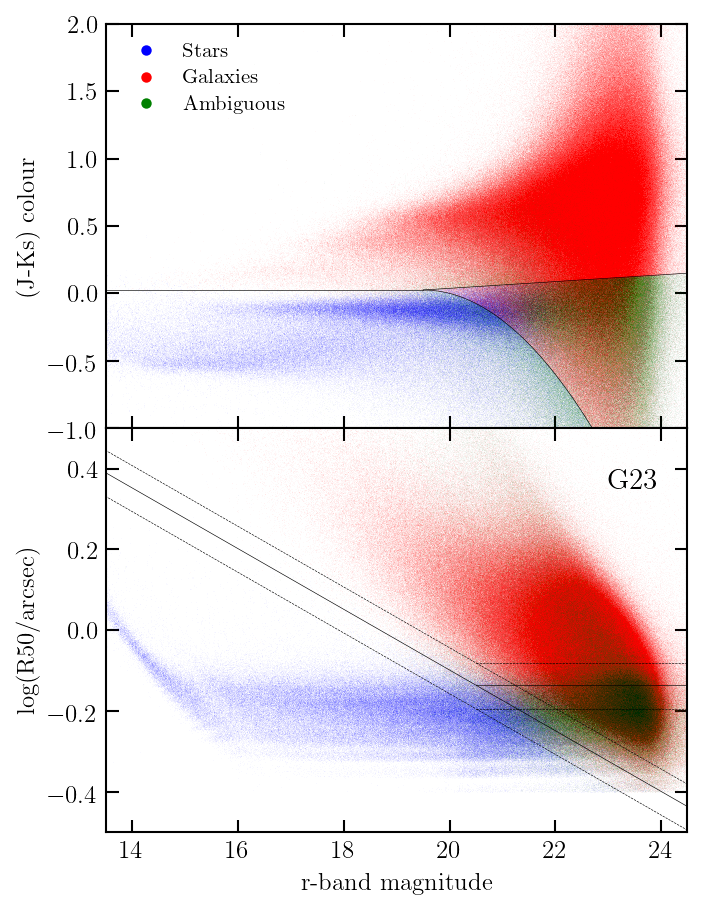}

\caption{Star-galaxy separation for the G23 field showing ($J-K_s$) colour versus magnitude (upper), and the measure half-light radius versus magnitude (lower) showing the stars, galaxies and ambiguous objects (in blue, red and green respectively). The solid lines denote the various cuts imposed (see text for full explanation), and the dashed lines show how these vary with seeing, with the 0.05-0.95 quantile range of the \textsc{log10seeing} value shown.  }
\label{fig:stargal}
\end{figure}

We initially plot $(J-K_s)$ v $r_t$ and $\log{R_{50}}$ v $r_t$ where $J$ and $K_s$ are the extinction-corrected colour measurements, $r_t$ is the extinction-corrected total $r$-band magnitude, and $R_{50}$ is the effective half-light radius of the dilated segment. The latter is determined from the number of pixels within the segment. We then draw two lines on each plot to define the galaxy regions, stellar regions, and the ambiguous regions (see the solid lines on Fig.~\ref{fig:stargal} that divide the data into three regions). 
If an object is the same class in both planes, then this class is adopted. If it is \texttt{ambiguous} in only one plane, then it gains the \texttt{galaxy}/\texttt{star} class, and if it is a \texttt{star} in one plane and a \texttt{galaxy} in the other, then it gains the \texttt{ambiguous} class. 
The equations used to separate the parameter spaces in $(J-K_s)$ v $r_t$ space are given by:
\begin{align}
\begin{split}
    (J-K_s) = 0.025, & \qquad\text{if } r_t < 19.5\\
    (J-K_s) = 0.025+0.025(r_t-19.5), & \qquad\text{if }r_t > 19.5\\
    (J-K_s) = 0.025-0.1(r_t-19.5)^2, & \qquad\text{if }r_t > 19.5, 
\end{split}
\end{align}
and in $\log(R_{50})$ v $r_t$ space:
\begin{align}
\begin{split}
    \log(R_{50}) = \Gamma+0.05-0.075(r_t-20.5), & \quad\text{any }r_t\\
    \log(R_{50}) = \Gamma+0.05, & \quad\text{if }r_t > 20.5, 
\end{split}
\end{align}
where $\Gamma$ is the median \textsc{log10seeing} value. 

Finally, based on the match to the GAMA redshift catalogue, we reassign any object with a confidently (\textsc{NQ} $> 2$) measured redshift above 0.002 to have a {\sc class} flag of \texttt{galaxy}, and any object with a quality measured redshift of $-0.002 < z < 0.002$ to have a {\sc class} flag set to \texttt{star}.

The process hence starts with ambiguity and refines the classifications through a staged process using colour and size, then redshifts. Fig.~\ref{fig:stargal} shows the detected objects in the G23 region, coloured by their final {\sc class} flag as indicated. Note that ambiguous objects are by definition those which reside in \textit{both} the ill-defined regions, unless a redshift is known or the object has been visually inspected.

\subsubsection{Cleaning spurious detections}
\begin{figure*}
\includegraphics[width=\textwidth]{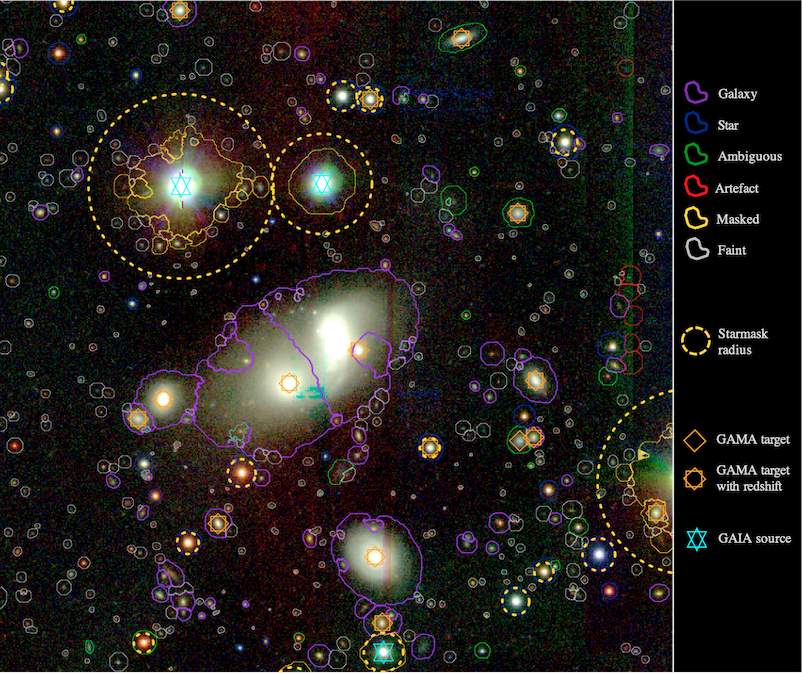}
\caption{An example 6 arcmin by 6 arcmin region showing the final classifications as described in the text: galaxies with $Z<21.5$ mag (purple), stars with $Z<21.5$ mag (blue), ambiguous with $Z<21.5$mag (green), artefacts (red), faint (grey) and masked (yellow). For masked stars, the starmask region is shown with a dashed yellow circle. GAMA targets are shown as orange diamonds, and GAIA sources are indicated with cyan stars. The region shows a number of complexes where the watershed deblending technique (by which every pixel is assigned to a single object only) is apparent.}
\label{fig:example}
\end{figure*}

\begin{figure}
\includegraphics[width=\columnwidth]{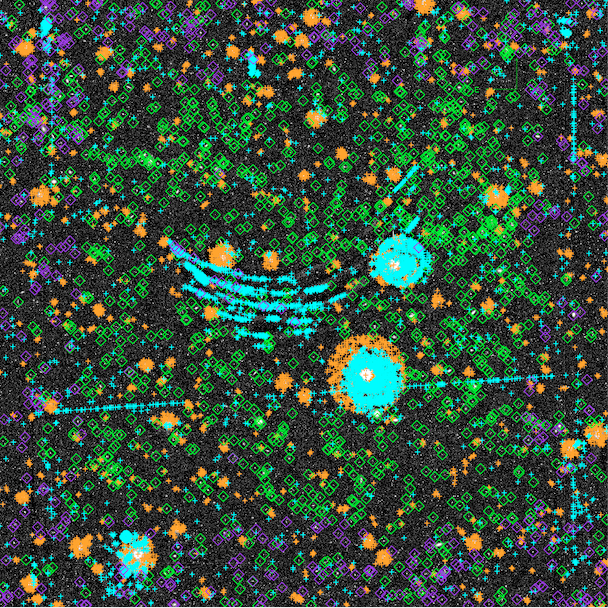}
\caption{Problematic sq degree region showing many of the issues we face. Overlain on the image are objects within starmask regions (orange) and objects labelled as artefacts (pale blue).
Other objects that are bright but have no known GAMA counterpart are shown in purple. Finally, known GAMA galaxies are shown in green.}
\label{fig:crudplot}
\end{figure}

As with any dataset the VST and VISTA imaging contains a variety of spurious detection issues, with origins varying from diffraction spikes and
offset ghosts, to baffling issues, noisy stacks, stack edge effects, transient objects (including Mars in one frame), and satellite trails etc. 
To identify artefacts we use a series of cuts to highlight objects in improbable parameter space. These have been arrived at through fairly extensive testing and visual checking and in particular viewing bright objects for which no match exists in the
GAMA catalogue. 
From this process we arrive at a series of diagnostic flags and cuts as indicated in Table \ref{tab:artefactDiagnostics}. Note that the table is progressed from top to bottom allowing overrides, hence we move from less certain to more certain classification
markers. 
Fig.~\ref{fig:crudplot} shows a single sq degree for a problematic region indicating some of the issues: bright star ghosting, baffling issues, missing data, offset ghosting and frame edge effects.
Overlain are objects with \textsc{starmask} flag set (orange), class set to artefact (cyan), and some remaining objects with bright fluxes not previously detected in the SDSS GAMA catalogue (purple). About half of the objects in this latter category represent new objects. 
An example of the object classifications is shown in Fig. \ref{fig:example}, where the segment of each object in this field is coloured by the corresponding classification, as indicated in the caption. 
The process is never going to be perfect but we believe the unclassified artefact rate is now well below 1 per cent of the galaxy population. 
We also note that no objects are removed and hence alternative or additional cleaning can be applied as we improve our understanding of the data.

\begin{table*}
\caption{A summary of the diagnostic markers, note diagnostics higher in the table
  supercede those lower down, hence the table also represents a
  priority when updating the same flag \label{tab:flags}}
\begin{tabular}{c|c|c|p{5.5cm}|p{5.5cm}} \hline
Flag & default & setting & criteria & reason \\ \hline \hline
class & ambiguous & galaxy & GAMA match with $Z > 0.002 \& NQ > 2$ & Known GAMA galaxy\\
class & ambiguous & star & GAMA match with $-0.002 Z < 0.002 \& NQ > 2 $ & Known GAMA star \\
class & ambiguous & artefact & $m_{\rm rt} - m_{\rm Zt}) < -0.75$ & Improbable $(r-Z)$ colour \\
class & ambiguous & artefact & $log10(R50) < -0.4$ & objects size is smaller than one pixel \\
class & ambiguous & artefact & No detection in two of $gri$ bands but optical data exists & Only detecetd in 1 optical band \\
class & ambiguous & artefact & No detection in three of $ZYJHK_s$ bands but near-IR data exists & Only detected in 1 or 2 near-IR bands \\
class & ambiguous & artefact & $\sigma_{\rm RMS} > median(\sigma_{\rm RMS})-5*{\rm st.dev.}(\sigma_{\rm RMS}$ & Elevated skynoise \\
class & ambiguous & star & {\rm starssize+starscol} $> 3.5 $ & star \\
class & ambiguous & galaxy & {\rm starssize+starscol} $< 0.5$ & galaxy \\ \hline
starmask & 0 & 1 &  $r[\mbox{arcmin}]=10^{(1.6-0.15g)}
\mbox{ and } [r < 5.0', g < 18.0]$ & object lies near a bright GAIA DR2 star\\ \hline
duplicate & 0 & 1 & Does not appear in all overlapping regions & Any object that hasn't been detected in all overlapping square degrees is spurious.  \\
duplicate & 0 & 1 & Duplicate identified by coordinate match with either \texttt{cen} or \texttt{max} coordinates, flag is assigned to the object with less flux & Segment is incomplete, either due to frame edge, or insufficient segment rebuilding.   \\\hline
mask & 0 & 1 & RA, Dec & Inside the GAMA footprint \\ \hline
Z\_confusionFlag & 0 & 1 & $\Delta z$ of multiple GAMA target matches $> 0.1$ & Segment contains flux from multiple GAMA objects at different redshifts.  \\ \hline
\label{tab:artefactDiagnostics}
\end{tabular}
\end{table*}

\subsection{Merging the {\sc ProFound} catalogues}
\label{sec:catMerging}

\begin{figure}
\includegraphics[width=0.475\textwidth]{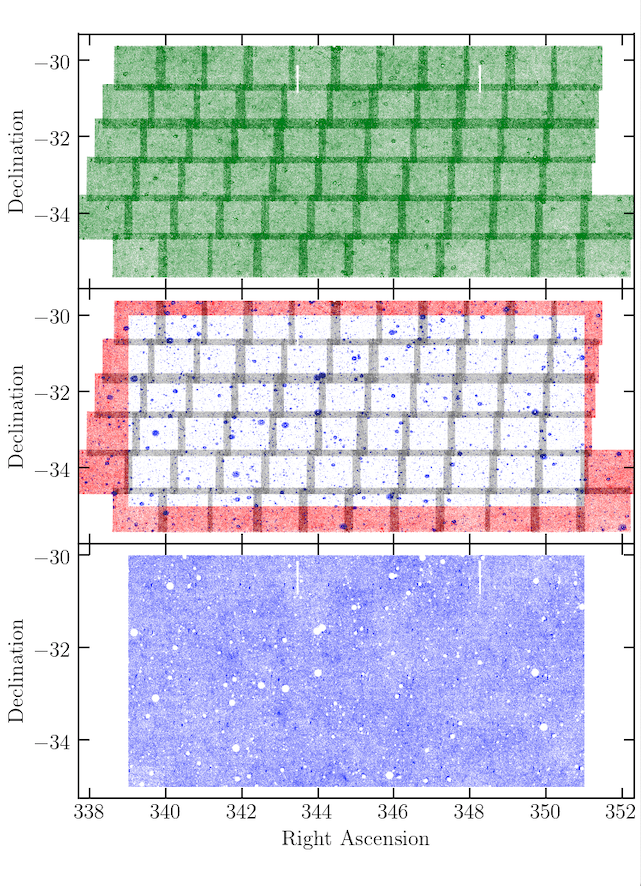}
\caption{The progression of the data flow for the G23 field. (top) {\sc ProFound} is used to
  process each sq degree {\sc tile} plus overlap to generate the star plus
  galaxy source catalogue shown (green points). (centre) the GAMA
  boundary is applied (red) and galaxies in the overlap region
  identified (grey), the star-mask is then applied (blue) and regions
  with missing photometry identified (cyan). (lower) the final galaxy
  catalogue (blue) showing only those galaxies in high quality regions
  away from bright stars or missing data.}
\label{fig:tiles}
\end{figure}

Having generated 280 independent sq degree tiles across the four GAMA fields, we now need to combine these into four catalogues (one for each GAMA region), by merging while removing duplicates in the overlap regions. 
Fig.~\ref{fig:tiles} highlights the issue by showing the concatenated {\sc tile} catalogues without consideration of duplicates. 
The over-density of the overlap regions manifests as a regular tartan-like pattern representing the double detection of objects in these regions.
A simple coordinate match in the overlapping regions is insufficient to identify duplicates, as there are a number of reasons why the two sets of coordinates may not be identical:
\begin{itemize}
    \item An object in the overlapping region may have been visually regrouped in two slightly different ways. If so, some segments may exist in one field, but not the other. 
    \item Very noisy segments, due to slightly different \textsc{ProFound} sky solutions, may have different boundaries and hence different coordinates. 
    \item Segments that include multiple sources (which can occur if the saddle point between sources falls above the \textsc{cliptol} threshhold) may also have different \textsc{RAmax} and \textsc{Decmax} values\footnote{For each object, \textsc{ProFound} computes two sets of coordinates. \textsc{RAmax/Decmax} values indicate the coordinate of the brightest pixel within a segment, whereas \textsc{RAcen/Deccen} values show the flux-weighted central coordinate of the segment.} if the sky solutions differ slightly. 
\end{itemize}

In all three of the above scenarios, a simple coordinate match will not identify ``lone" objects that have not been exactly duplicated, and these objects will be counted twice. The following measures have been taken to account for these scenarios:
\begin{itemize}
    \item A check for duplicate objects is conducted exclusively in the overlapping regions. In these regions, any object that is \textit{not} duplicated is immediately assumed to be spurious, and is assigned {\sc duplicate}=1. In the scenario of differently merged objects, this ensures that only the ``main" segment of the galaxy will be considered. Additionally, all noisy segments without stable coordinates will be given {\sc duplicate}=1.  
    \item For each successfully duplicated object, priority is given to the duplicate with more flux. In the scenario of differently merged objects, this ensures that only the most aggressively regrouped version of an object will have {\sc duplicate}=0. Additionally, if objects have been broken up due to proximity with the edge of the tile, then the version of the object that is most complete will be prioritised. 
    \item In the overlapping regions, a duplication is checked for with both the \textsc{RAmax}/\textsc{Decmax} coordinates, and also the \textsc{RAcen}/\textsc{Deccen} coordinates. If an object is flagged as being a duplicate in either one, then it will be assigned a value of {\sc duplicate}=1. This has been done because even if a stable segment has differing \textsc{RAmax}/\textsc{Decmax} values due to a slightly different sky solution, the \textsc{RAcen}/\textsc{Deccen} values tend to be the same. As such, segments whose brightest pixel position fluctuate due to different sky solutions are adequately accounted for. 
\end{itemize}
Hence, {\sc duplicate}=0 will produce a catalogue of unique objects, whereas {\sc duplicate}=1 will produce a catalogue of redundant objects (see the blue-grey bands in the centre panel of Fig \ref{fig:tiles}). 
Note than no objects are removed from the catalogue at any stage, instead we introduce flags to enable extraction  of well-defined samples.

\section{Extracting photometry in the mid- and far-IR}
\label{sec:FIR}

Objects in the MIR-FIR are unresolved, unlike the FUV-MIR bands where objects are either fully or partially resolved.  
In addition, some of the brightest far-IR sources may have no or minimal optical fluxes, and vice-versa. 
Finally, the depth of the FIR imaging is also lower than optical images (as shown in Fig. \ref{fig:g23depth}). 
We therefore run \textsc{ProFound} in a manner that utilises a different measurement technique in the MIR-FIR.
To account for the above differences between the MIR-FIR and FUV-MIR, the FIR fluxes of optically-identified objects are iterated over by applying expectation maximisation (EM), as we describe in the following section. 

The philosophy of this measurement approach is to model the FIR flux of each optically-detected object, and then iterate over the flux of the object in each band, ensuring that all FIR flux is accounted for. 
In each band separately, the locations of optically-selected objects provide the coordinates at which to fit for objects. 
For the W3-W4 bands, the input objects are selected to be all objects from the optical catalogue with $m_r<20.5$ mag,  excluding those objects with an \textsc{uberclass}$=$\texttt{artefact} flag. 
Because stars are still bright at these wavelengths, we make sure to model their flux contribution. 
In the $P100$-$S500$ bands, however, we also remove all objects with \textsc{uberclass}$=$\texttt{star} or \texttt{ambiguous} flag, as these objects are not expected to emit FIR flux. 
A magnitude guess for each object is determined by running an initial round \textsc{ProFound} at the input coordinates.
Based on this initial magnitude guess, \textsc{ProFit}\footnote{Available on Github: \url{https://github.com/ICRAR/ProFit}} \citep{Robotham17} is applied to construct a model of each object given the PSF of each band and the magnitude guess of each object, to create a model of the FIR-emitting sources within the tile. 
This \textsc{ProFit} model is then subtracted from the image to produce a residual image. Iteratively, the model fluxes are modified using expectation maximisation in order to minimise the flux remaining in the residual image. 
After these initial iterations, any additional FIR-emitting sources that are not present in the original optical catalogue will be apparent. 
These are now identified using \textsc{ProFound} in source detection mode, and are included from this point onwards in the analysis as ``additional sources". 
The inclusion of these additional sources mainly serves the purpose of accurate flux determination, to ensure that no background FIR flux is attributed to a foreground catalogue object\footnote{Note that if a background source exists within the PSF of the foreground object, that this flux will no longer be separated. Such a scenario was presented by Allison et al. (submitted), where the SED of the target foreground object was likely polluted in the FIR through the contribution of a high-$z$ background object.}.
Once a catalogue of both optically-selected and additional sources has been determined, \textsc{ProFound} is run over the full image to determine the magnitudes of all objects, again in an iterative sense. Some of these objects will be marginalised out by \textsc{ProFound}, producing very low flux values, with correspondingly large uncertainties. 
This process is conducted using the \texttt{profoundFitMagPSF} command, which we show for completeness in the Appendix. 

To account for potential oversubtraction of the sky in the data reduction phase of the FIR imaging (possible due to the confused nature of sky pixels), we do an explicit sky subtraction in each band in a second phase. During the detection phase of additional sources, as described above, \textsc{ProFound} makes a measurement of the sky. In this second phase, we rerun the command as described above, in which the input image has undergone an explicit sky subtraction. 

In total over the four GAMA fields, photometry was measured for 822,326 objects with an \textsc{uberclass}=\texttt{galaxy} flag with $m_r<20.5$ mag. The fraction of these objects with detections in each band (defining a detection as having a measured magnitude $<30$) varied in each band, depending on the depth of the imaging. In the $W3$/$W4$/$P100$/$P160$/$S250$/$S350$/$S500$ bands, 62/42/38/43/50/42/32 per cent of objects had detections, respectively. 38,413 objects had detections in all seven bands, corresponding to just under 5 per cent of the $m_r<20.5$ mag galaxy sample. 

\begin{figure}
\includegraphics[width=\columnwidth]{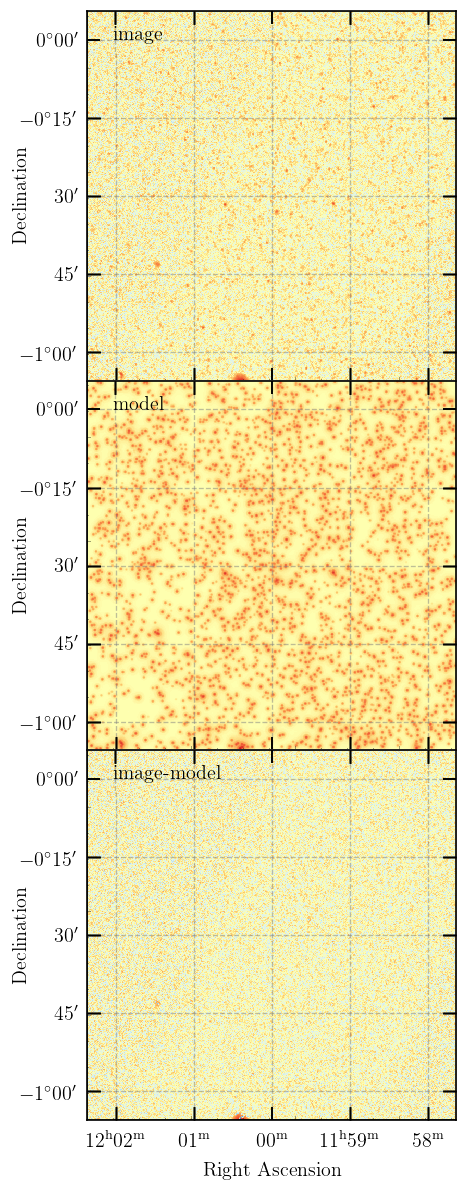}
\caption{Example of the photometry extraction in the SPIRE 250 band for a single tile in the G09 region. The top panel shows the original image, and the middle panel shows the \textsc{ProFit}-produced model of the objects. The residual when subtracting the model off the image is shown in the bottom panel. }
\label{fig:FIRphotometryresidual}
\end{figure}

The resulting uncertainties, due to the EM mixture modelling process, accurately reflect the inherent uncertainties in this process. If two optically-detected sources are close in projection, then the uncertainty will reflect the potential confusion between these two sources.   
We show in Fig. \ref{fig:FIRphotometryresidual} an example of the original image, final modelled image, and resulting residual of a specific square degree. 
Note that in this example, the main features in the residual image originate from resolved galaxies that are not well-modelled by the PSF. 
In addition to the objects detected in the FIR with optical counterparts, $4.5/43.4/0.3/0.4/23.6/19.6/4.9$ per cent of objects in the $W3$/$W4$/$P100$/$P160$/$S250$/$S350$/$S500$ bands were identified as ``additional sources". 
Generally this percentage is reflective of the depth of the imaging. We note that the large number of additional objects detected in the $W4$ band indicates that the \textsc{ProFound} parameter used to identify these objects was likely too aggressive in this band.

The above process is conducted independently for each band in each individual tile. In total, it requires less than one hour to extract the W3-S500 photometry for a single square degree. 
The outputs for the additional sources are saved, however are not included in our final catalogue. An analysis of these sources is beyond the scope of this work. 

\section{Verification and validation}
\label{sec:verification}

\subsection{Comparison with previous photometry}

\subsubsection{Comparison to SDSS/2MASS}

\begin{figure}
 \includegraphics[width=85mm]{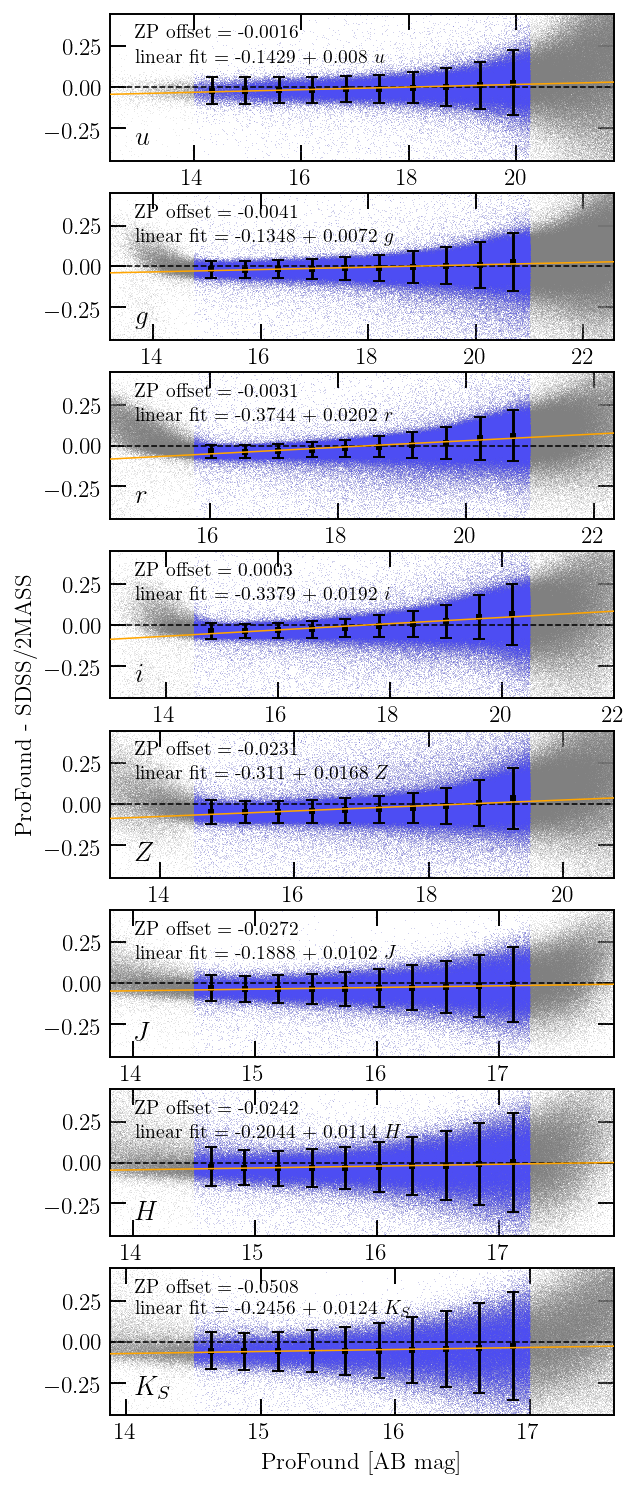}
 \caption{Comparison of the new {\sc ProFound} photometry versus the SDSS/2MASS photometry, for the combined G09+G12+G15 data in the filters indicated for stars only. In each panel we show the data (grey dots for the full sample, and blue dots within the fitted magnitude range), the median offset (black points with 1$\sigma$ range), and the linear fit to the offset (orange line).  The offset and the least squares fit to the median values are shown in the upper left of each panel.}
 \label{fig:photom}
 \end{figure}

Fig.~\ref{fig:photom} shows a comparison of the {\sc ProFound} measured photometry of stars (in selected magnitude ranges) to either SDSS DR13 ($ugriz$) or 2MASS ($JHK_S$). For the comparison to SDSS in the $u$ and $r$ bands we use the filter conversions as outlined by \citet{Kuijken19}. For the comparison to 2MASS, we include the appropriate corrections for the 2MASS filter shapes, colour terms, and AB to Vega corrections given in \citet{Gonzalez-Fernandez18}. 
 The effect of saturation in the VST $gri$ and VISTA $Z$ bands is evident at the bright end, where \textsc{ProFound} recovers less flux for the bright stars than was measured by SDSS. 	
 Note that the SDSS data are \emph{not} corrected for the known $u$ and $z$ offsets from the SDSS system to the AB system. 
 In all bands the photometry agrees well with scatter about the equality line increasing slightly towards fainter fluxes.
 In all cases the offsets are $<0.06$ mag with the largest offset seen in the $K_S$ band. 

\subsubsection{Comparison to LAMBDAR}

\begin{figure}
\includegraphics[width=\columnwidth]{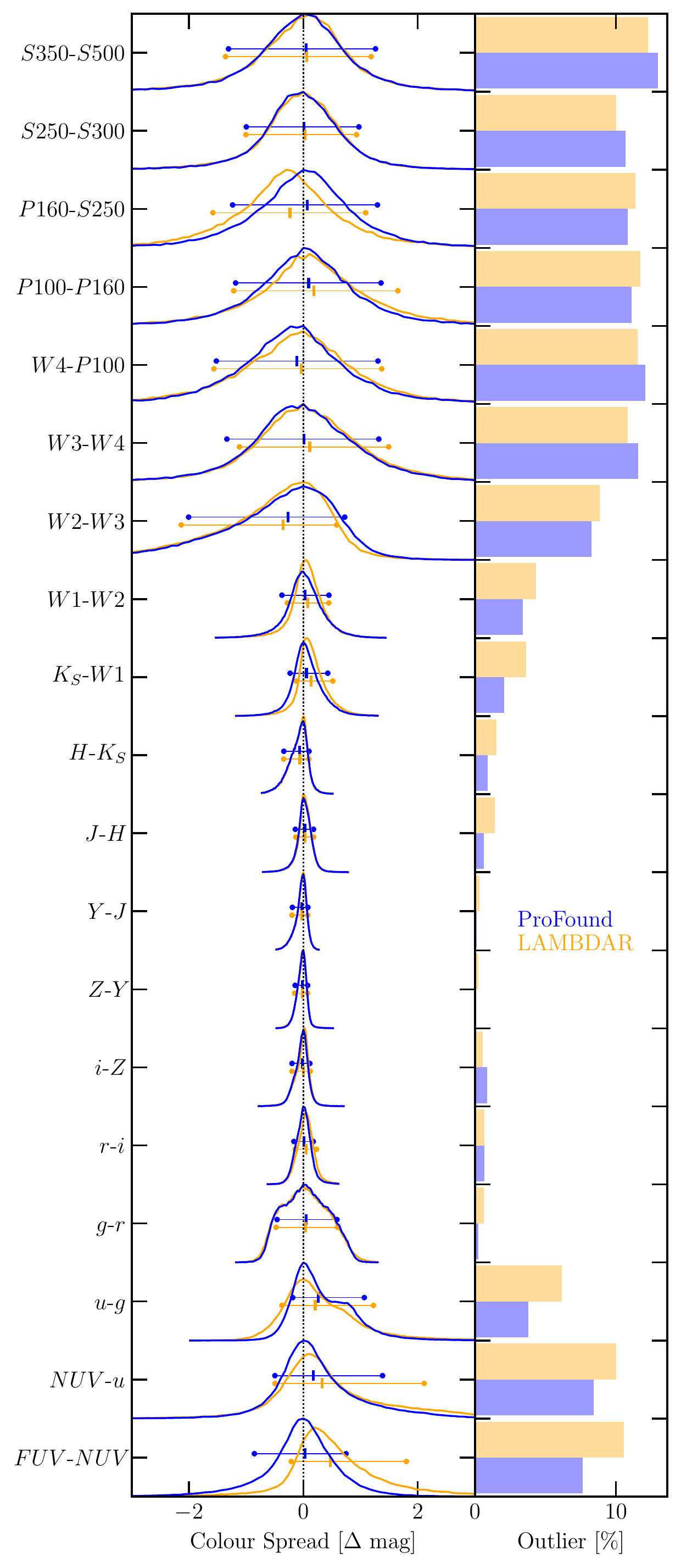}
\caption{Colour distributions of all adjacent bands for the photometry presented in this paper (blue), as compared with LAMBDAR \citep{Wright16} (orange) for all matching galaxies. Individual colour distributions are shown in the left-hand panel, relative to the peak of the \textsc{ProFound} colour distribution. We show the 0.1-0.9 quantile ranges for each sample, as well as the median value. The right-hand panel shows the outlier percentage for each sample, quantified as the percentage of points that lie further than 0.5 mag away from the 0.1-0.9 quartile range. }
\label{fig:ColourDist}
\end{figure}

We compare the colour distributions of the photometry presented in this paper against equivalent colour distributions produced by LAMBDAR \citep{Wright16} for the subset of objects with measured 20-band photometry, as shown in Fig. \ref{fig:ColourDist}. 
The left panel of the plot features the relative colour distribution of the overlapping sample between LAMBDAR and the new photometry. 
Blue histograms refer to the new photometry presented in this paper, whereas mustard histograms refer to the LAMBDAR photometry. For each histogram, we show the 0.1-0.9 quartile range with a horizontal line, where the median value is shown. 
Note that the colour values have been shifted so that the peak in the distribution for the new photometry is at 0. 
The right panel of the plot shows the number of outliers in each sample for each band, where outliers are determined to be objects that have colours more than 0.5 mag outside the 0.1-0.9 quartile range. 
The new colour distributions are improved in the UV, optical and also (marginally) the FIR bands, however we note that the colour distributions in the NIR bands are better in LAMBDAR. This is likely the result of the generous segment dilation we have implemented in order to catch halo flux, adding more sky noise than in LAMBDAR. 

\subsection{Seeing variations across tiles}

By analysing the \textsc{R50} values of stars across the four GAMA fields, we are able to determine the extent to which the seeing varies across the tiles in each of the optical/NIR bands. The median \textsc{R50} value varies between $\sim$ 0.4 -- 1 arcseconds over the four fields, in each of the different photometric bands. 
Due to the nature of the segment dilation within \textsc{ProFound}, the total flux within each object is accounted for despite potential variations in the PSF. To check that these PSF variations are not having an impact on the derived fluxes for galaxies in the sample, we assess how the minimum \textsc{R100} value for galaxies compares to the PSF in each tile. For galaxies down to an $r$-band magnitude of 23, we find that the smallest galaxies are on average larger than the PSF by factors of 2 -- 4. This confirms that galaxy segments are consistently larger than the PSF, meaning that seeing variations across the tiles are not affecting the measured fluxes of galaxies in our photometry. 

\subsection{Final astrometric accuracy}

\begin{figure}
\includegraphics[width=\columnwidth]{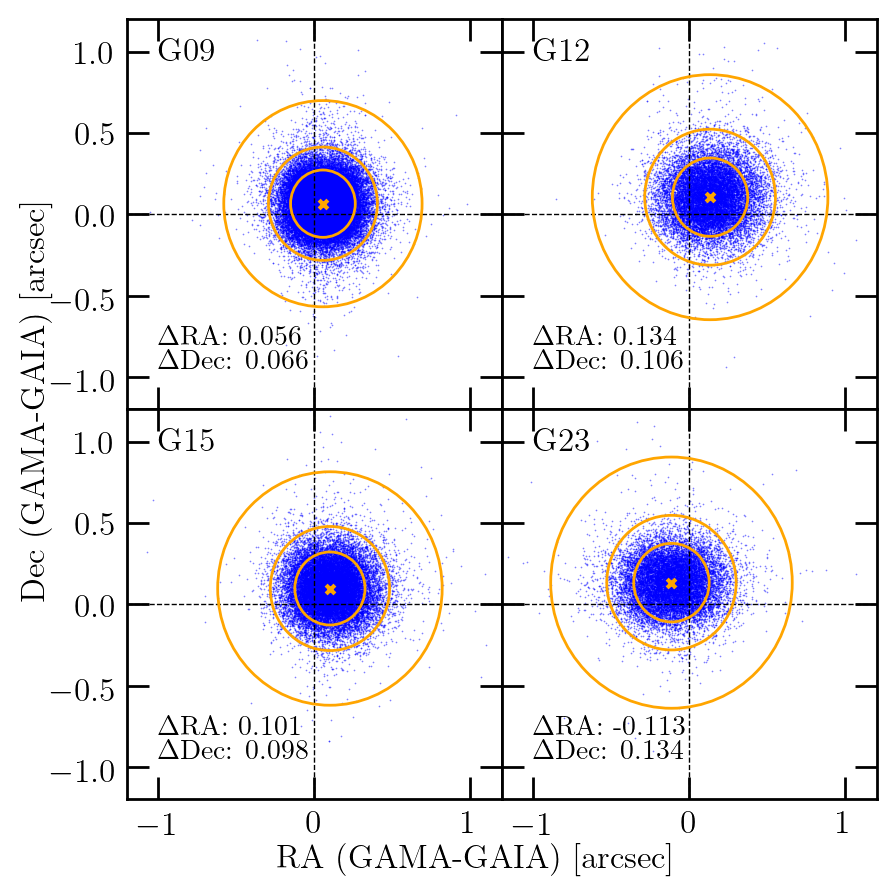}
\caption{Astrometric accuracy of the native VST KiDS data against GAIA DR2 for each of the GAMA regions as indicated.}
\label{fig:astrom}
\end{figure}

To verify the astrometric accuracy we identify objects selected as stars in the range $ 14 < m_r < 16$ and match against GAIA DR2 \citep{Gaia18, Lindegren18} taking the best match within 3 arcsec. Fig.~\ref{fig:astrom} shows the RA and Dec offset for each field, with the medians indicated (orange crosses) and the circles enclosing 1, 2, and 3$\sigma$. 
In all cases the 3$\sigma$ astrometry error is within $\pm0.75$ arcsec but some modest offset is seen. We therefore add two extra columns {\sc RAGAIA} and {\sc DecGAIA} where we correct the VST KiDS astrometry to the GAIA astrometric frame by implementing a simple RA and Dec offset appropriate to each region. The offsets are as shown in Table.~\ref{tab:areas} (columns 7 and 8).

\subsection{Star and galaxy counts in each region}
\label{sec:counts}
\begin{figure*}
\includegraphics[width=\textwidth]{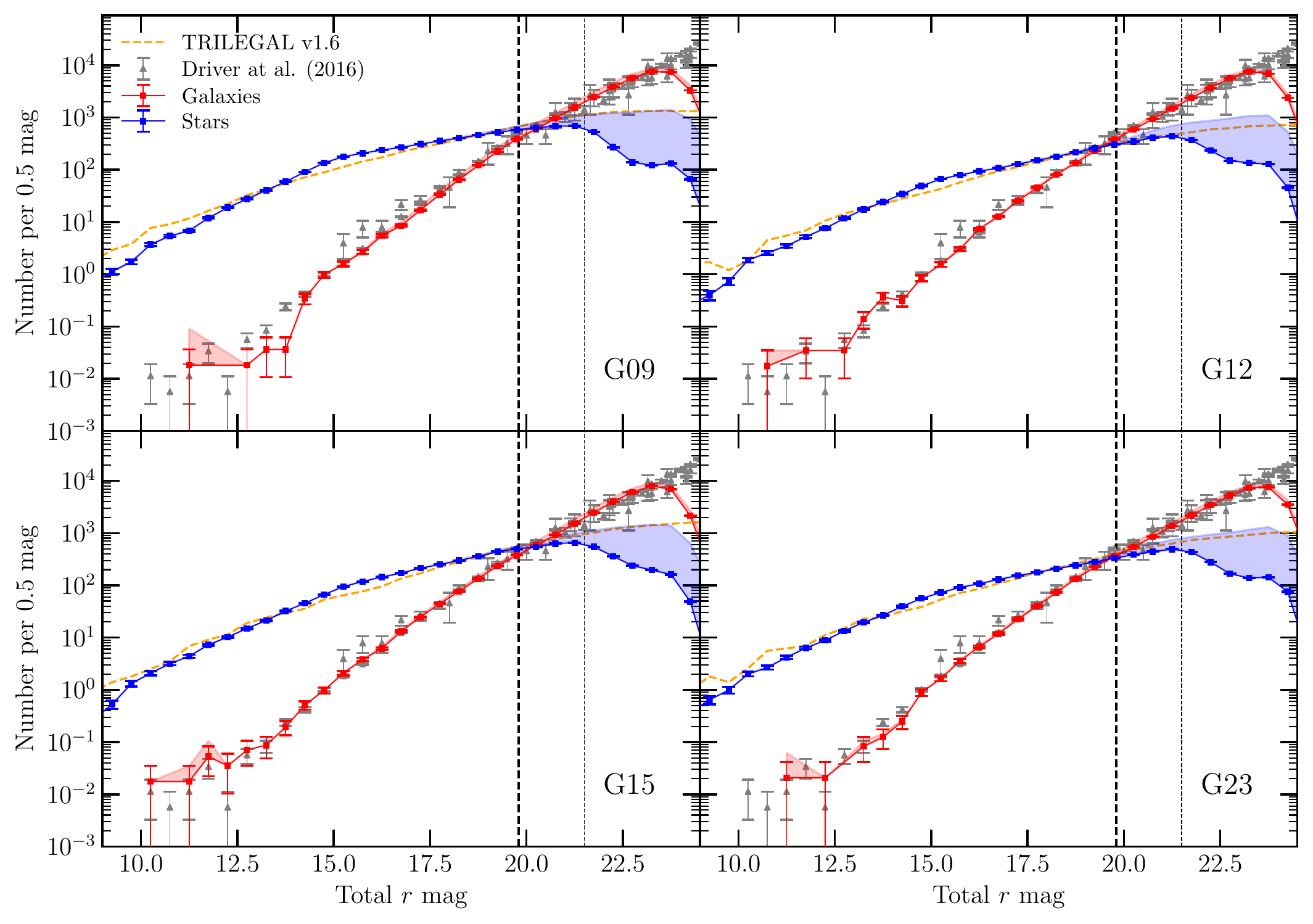}
\caption{Galaxy and star counts for each of the four GAMA fields, as labelled. The errors are the combination of root-n statistics (shown as errorbars, dominating at bright magnitudes) combined linearly with the star-galaxy classification error (shown as the coloured region, dominating at faint magnitudes). The vertical lines show the nominal GAMA spectroscopic survey limit (dashed lines) and the upcoming WAVES spectroscopic survey limit (dotted line). Data shown are taken from the compendium of \citet{Driver16c} that combined ground-based and space-based filters through comparable, but not identical, bandpasses.
\label{fig:ncounts}}
\end{figure*}

As an initial diagnostic we construct the galaxy number-counts for each region and compare to literature data \cite{Driver18} in the $r$ band (see Fig. \ref{fig:ncounts}). 
Within each panel we show the galaxy counts (red) and the star-counts (blue), where the shaded regions indicate the range covered when ambiguous objects are included. Note that the ambiguous objects are not explicitly included in either main line. 
Literature counts are shown in grey and the predicted star-counts from the \textsc{TRILEGAL} v1.6 model \citep[see][]{Girardi12}, are shown with an orange dotted line. 
In all four regions the recovered $r$ band counts show broad agreement with the literature data. The galaxy number counts are slightly below the literature counts at the bright end in both the G09 and G23 fields.
We note that the \textsc{TRILEGAL} models also agree well with our star-count data and suggest that the majority of
the ambiguous detections are likely stars (as the upper bound of the star counts most closely matches the \textsc{TRILEGAL} prediction), and as also suggested by Fig. \ref{fig:stargal}. More details will be discussed in Koushan et al. (in prep). 

\subsection{Revised panchromatic depth}

\begin{figure*}
\includegraphics[width=1.0\textwidth]{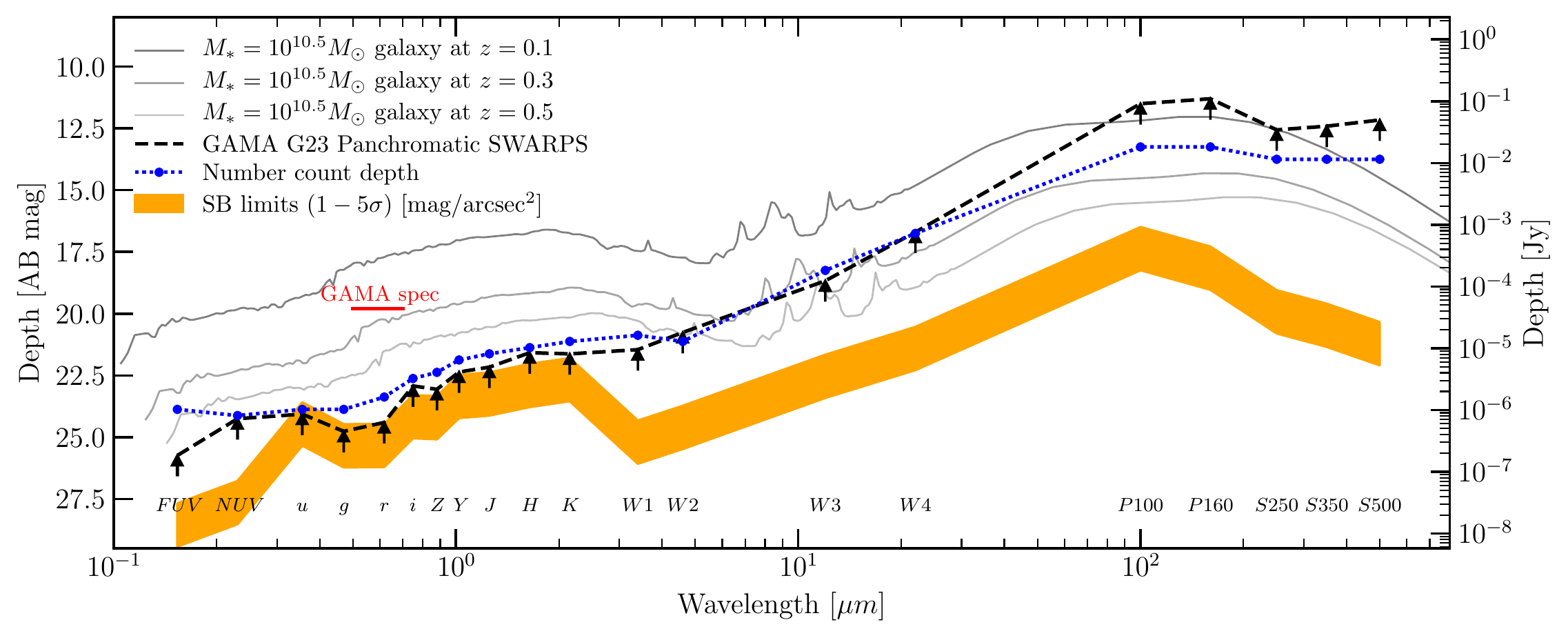}
\caption{The revised panchromatic depth for the G23 field, now including the VST KiDS data. The stated 5$\sigma$ depth levels in each of the bands is shown in black, whilst our depth measurement based on the turnover magnitude from the number counts in each band is shown as the blue dotted line. Not that this measurement is shallower in the FUV, $g$, $r$ and $i$ bands, while being deeper in the Herschel bands. The surface brightness limit of the images as determined by the \textsc{ProFound}-measured skyRMS values is shown in the yellow shaded region. The spectroscopic limit of the GAMA survey is shown in red. Template SEDs for a $M_*=10^{10.5} \rm{M}_{\odot}$ galaxy at varying redshifts are shown in grey. These are generated using the tool \textsc{ProSpect} \citep{Robotham20}. }
\label{fig:g23depth}
\end{figure*}

We show in Fig. \ref{fig:g23depth} the revised depth of our imaging and catalogues in each of the 20 photometric bands. In addition to showing the stated 5$\sigma$ depth limits for each of the imaging bands in black, we make two separate measurements of the depth of our data. The first utilises the number counts of each band, as described in the above Section \ref{sec:counts}, to estimate the depth in each band as the magnitude at which the galaxy counts turn over (calculated as the magnitude at which the counts are lower than the counts in the previous magnitude bin). This measurement is shown in Fig. \ref{fig:g23depth} as the dotted blue line. We note that this measurement is very close to the stated depths, except in the FUV, $g$, $r$ and $i$ bands where our depth measurement is shallower, and the Herschel bands in which our measurement is deeper. 

As an alternative way of indicating the image depth, we plot the surface brightness limits in each band as determined by the \textsc{ProFound}-measured skyRMS values in each band. These limits are shown in Fig. \ref{fig:g23depth} as the yellow shaded region. 

We compare these revised depths with a \textsc{ProSpect}-derived SED for a $M_*=10^{10.5}\rm{M}_{\odot}$ galaxy with a constant SFH at redshifts of $z=$ 0.1, 0.3, and 0.5. This comparison highlights that, with our sensitivity, but $z=0.3$ we are no longer sensitive to a galaxy of this stellar mass in the FIR.  

\subsection{Impact on existing GAMA studies}

The main impact that this new catalogue will have on existing studies comes from the new segmentation maps. As discussed in Section \ref{sec:GAMAmatch}, not all objects in previous GAMA studies will be listed in this new catalogue, as they have been merged with neighbouring GAMA objects, or because the target was based on an artefact in SDSS imaging. This is the case for $\sim$ 4 per cent of GAMA targets.  
As a test of the impact of this on existing GAMA studies, we assess the number of objects in existing data products that no longer appear in the new catalogue. 
In the catalogue of visual morphologues (DMU \textit{VisualMorphologyv3} \citealt{Kelvin14}, as used in, for example, \citealt{Kelvin14b, Moffett16, Lange16, Moffett16b, Alpaslan15}), 0.81\% of objects are not in the new catalogue. For the group catalogue (DMU \textit{G3Galv10} \citealt{Robotham11}, as used in for example \citealt{Alpaslan12, LaraLopez13, Alpaslan14, Robotham14, Davies15, Deeley17}), this number is only 0.26\%. For the SED-fitting catalogue using \textsc{MagPhys} (DMU \textit{MagPhysv06} \citealt{Driver16}, as used in, for example, \citealt{Davies17, Mahajan18, Driver18}) only 0.25\% of objects are missing, and for the catalogue of S\'{e}rsic indices (DMU \textit{SersicCatSDSSv09} \citealt{Kelvin12}, as used in, for example, \citealt{Kelvin14, Deeley17, Bremer18}) 0.87\% of objects are missing. 
This corresponds to an absolute number of 314/470/487/1935 galaxies respectively. 
We highlight that this number is very small, and hence do not expect that this will have any impact on existing GAMA studies. 
The \textsc{GAMAKidsVikingCATAIDMatchv01} catalogue will aid in the bookkeeping of any such circumstances, by identifying for each GAMA target the corresponding \textsc{uberID} in the new catalogue. 

\section{Access to catalogues}
\label{sec:catAccess}

The catalogues are available to any member of the public via a collaboration request\footnote{\url{http://www.gama-survey.org/collaborate/}}.
Included in the release are the Data Management Units (DMUs) as described in Table \ref{tab:DMUs}. 

\begin{table*}
\caption{Data Management Units (DMUs) released as part of this paper.}
\begin{tabular}{r|p{8.0cm}} \hline \hline
DMU & Description \\ \hline
GAMAKidsVikingv01 & \textsc{ProFound} FUV-NIR analysis described here.  \\
GAMAKidsVikingCATAIDMatchv01 & CATAID match of legacy catalogues to new catalogue.  \\
GAMAKidsVikingFIRv01 & \textsc{ProFound} FIR PSF-limited analysis described here.  \\
GAIADR2GAMAstarmaskv01 & List of GAIA DR2 stars that define the starmask. \\  \hline
\end{tabular}
\end{table*}
\label{tab:DMUs}

Each of these include DMU descriptions in the GAMA Schema Browser and tools to access the data via either the Single Object Viewer \citep{Liske15} or the Panchromatic SWARP Imager \citep{Driver16}. Note that the spectroscopic component is described in full in \cite{Liske15}. Note that version numbers may change as products are updated however older versions are available via the Schema Browser.

\subsection{Usage of DMUs}
In the analysis described within this paper, we have kept all measurements and introduced a series of flags (see Table.~\ref{tab:flags}) to indicate various issues. Hence, the extraction of a catalogue suitable for science requires the use of these flags. 
For example, to extract all galaxies in the GAMA G09 region outside the starmask with robust spectroscopic redshifts from the \textsc{KidsVikingGAMAv01} catalogue one must execute the following:
 
\texttt{uberclass="galaxy" \& duplicate=0 \& mask=0 \& starmask=0 \& region="G09" \& NQ $>$ 2}
 
This results in 59930 galaxies covering 54.93 sq degress of sky (see Table \ref{tab:areas}).

The FIR catalogue (\textsc{KidsVikingGAMAFIRv01}) only includes a subset of the objects in \textsc{KidsVikingGAMAv01}, as per the description in Section \ref{sec:FIR}. In all other respects, however, the structure of the catalogue is the same. The same set of flags to isolate catalogues has been implemented. Conducting the above command on this catalogue hence results in 58967 galaxies covering 54.93 sq degress of sky (see Table \ref{tab:areas}).

\section{Summary}
\label{sec:summary}

We present in this paper updated photometry of four GAMA fields in 20 wavelength bands, using deeper imaging from KiDS/VIKING and the code \textsc{ProFound}. 
\textsc{ProFound} has been separately run in first a resolved mode to extract photometry from first the $FUV$-$W2$ bands, and then in an unresolved to extract photometric measurements from the $W3$-$S500$ bands. 
In the resolved mode, sources are detected, fragmented sources are visually rebuilt, and then the corresponding fluxes in all bands are extracted, before correcting these values for dust exinction, defining star masks, and classifying the individual objects into classes of \texttt{star}, \texttt{galaxy}, or \texttt{ambiguous}. Spurious objects in the catalogue are assigned the class of \texttt{artefact}. 
The photometry in the unresolved regime is conducted using a subset of the resolved catalogue as an input, where fluxes are extracted in a Bayesian manner for objects that are identified as galaxies with $m_r < 20.5$. As part of this process, we identify additional sources that contribute flux in each FIR band (most probably background sources), however we leave an analysis of these objects to future work. 

As a verification of the new photometry, we have checked the astrometry of stars in our catalogue against their coordinates in GAIA, to identify our astrometric accuracy. 
A comparison of the galaxy and star number counts to literature data from \citet{Driver16} show that our galaxy number counts are consistent with the literature in all fields, with only a slightly smaller count number at the bright end in the G09 and G23 fields. 
We identify that the colour distributions in our new photometry, as compared with the previous LAMBDAR photometry \citep{Wright16}, are significantly tighter in the UV and optical bands, and also marginally better in the FIR bands. 

In future work (Bellstedt et al. in prep.), this photometry will be used to conduct SED-fitting with the newly-developed code \textsc{ProSpect} \citep{Robotham20} to measure stellar masses, star formation rates and star formation histories for individual galaxies.

\section*{Acknowledgments}

We thank the anonymous referee for their careful reading of this work, and whose comments improved the paper. 
SPD and SB acknowledge support from the Australian Research Council under Discovery Project Discovery 180103740.

GAMA is a joint European-Australasian project based around a spectroscopic campaign using the Anglo-Australian Telescope. 
The GAMA input catalogue is based on data taken from the Sloan Digital Sky Survey and the UKIRT Infrared Deep Sky Survey. 
Complementary imaging of the GAMA regions is being obtained by a number of independent survey programmes including GALEX MIS, VST KiDS, VISTA VIKING, WISE, Herschel-ATLAS, GMRT and ASKAP providing UV to radio coverage. 
GAMA is funded by the STFC (UK), the ARC (Australia), the AAO, and the participating institutions. The GAMA website is \url{http://www.gama-survey.org/}.
This work is based on based on observations made with ESO Telescopes at the La Silla Paranal Observatory under programme IDs 177.A-3016, 177.A-3017, 177.A-3018 and 179.A-2004, and on data products produced by the KiDS consortium. The KiDS production team acknowledges support from: Deutsche Forschungsgemeinschaft, ERC, NOVA and NWO-M grants; Target; the University of Padova, and the University Federico II (Naples).

This work was supported by resources provided by the Pawsey Supercomputing Centre with funding from the Australian Government and the Government of Western Australia.
We have used \textsc{R} \citep{R} and \textsc{python} for our data analysis, and acknowledge the use of \textsc{Matplotlib} \citep{Hunter07} for the generation of plots in this paper. 

\bibliographystyle{mnras}
\bibliography{BibLibrary}

\appendix

\section{ProFound commands}
\label{app:commands}

Here we show the command used to run \textsc{ProFound} in the multiband mode for the detection phase:

\vspace{0.3cm}
\noindent
\texttt{profoundMultiBand( }

\texttt{dir=}input directory,

\texttt{skycut=2.0,}

\texttt{pixcut=13, }

\texttt{ext=1,}

\texttt{tolerance=15,}

\texttt{reltol=-10,}

\texttt{cliptol=100,}

\texttt{detectbands=c(`r',`Z'),}

\texttt{multibands=c(`r',`Z'), }

\texttt{keepsegims=TRUE,}

\texttt{magzero=c(0,30),}

\texttt{dotot=FALSE, }

\texttt{docol=FALSE, }

\texttt{dogrp=FALSE,}

\texttt{verbose=TRUE,}

\texttt{boxiters=4,}

\texttt{grid=c(50,50),}

\texttt{roughpedestal=TRUE,}

\texttt{stats=FALSE,}

\texttt{groupstats=TRUE,}

\texttt{mask=0,}

\texttt{app\_diam=1.4,}

\texttt{fluxtype=`Jansky'}
\noindent
\texttt{)}

\vspace{0.5cm}
We also show the multiband call that was used to run the measurement phase after the regrouping of segments:

\vspace{0.3cm}
\noindent
\texttt{profoundMultiBand(}

\texttt{segim=fixed\_segim\footnote{\texttt{fixed\_segim} is the segmentation map produced using \texttt{profoundSegimKeep} after manually regrouping any fragmented segments.},}

\texttt{dir=}input directory,

\texttt{iters\_det=6,}

\texttt{iters\_tot=c(3,3,2,2,2,2,2,2,2,2,2,3,3),}

\texttt{totappend=`t',}

\texttt{sizes\_tot=c(15,15,5,5,5,5,5,5,5,5,5,15,15),}

\texttt{colappend=`c',}

\texttt{detectbands=c(`r',`Z'),}

\texttt{multibands=c(`FUV',`NUV',`u',`g',`r',`i',`Z',}
\texttt{`Y',`J',`H',`K',`W1',`W2'),}

\texttt{keepsegims=TRUE,}

\texttt{magzero=c(18.82,20.08,0,0,0,0,30,30,30,30,30,}
\texttt{23.16,22.82),}

\texttt{dotot=TRUE,}

\texttt{docol=TRUE,}

\texttt{dogrp=TRUE,}

\texttt{verbose=TRUE,}

\texttt{box=c(200,200,100,100,100,100,100,100,100,100,}
\texttt{100,200,200),}

\texttt{boxiters=4,}

\texttt{boxadd=c(50,50,50,50,50,50,50,50,50,50,50,50,}
\texttt{50),}

\texttt{grid=c(50,50,50,50,50,50,50,50,50,50,50,50,50),}

\texttt{roughpedestal=TRUE,}

\texttt{redosegim=FALSE,}

\texttt{deblend=FALSE,}

\texttt{groupstats=TRUE,}

\texttt{mask=0,}

\texttt{SBdilate=1.0,}

\texttt{SBN100=100,}

\texttt{app\_diam=1.4, }

\texttt{fluxtype=`Jansky'}
\noindent
\texttt{)}

\vspace{0.5cm}
The following commands presents the manner in which we have run proFound in the unresolved MIR-FIR regime, for each band separately:

\vspace{0.3cm}
\noindent
\texttt{profoundFitMagPSF(}

\texttt{RAcen=}RA coordinates, 
	
\texttt{Deccen=}Dec coordinates,
	
\texttt{image=}sky-subtracted image, 

\texttt{header=}image header, 
    
\texttt{psf=}band PSF, 
    
\texttt{magzero=}23.24/19.6/8.9/8.9/11.68/11.67/11.62\footnote{Zero-points as stated in table 3 of \citet{Driver16}.},
    
\texttt{magdiff = 5, }
    
\texttt{fit\_iters=5, }
    
\texttt{verbose = TRUE, }
    
\texttt{fluxtype=`Jansky',}
    
\texttt{doProFound=TRUE,}
    
\texttt{findextra=TRUE, }
    
\texttt{itersub=TRUE,}
    
\texttt{pixcut=3, }
    
\texttt{skycut=2, }
    
\texttt{ext=1, }
    
\texttt{redosky=FALSE,}
    
\texttt{iters=4, }
    
\texttt{tolerance=0, }
    
\texttt{sigma=}2/2/2/2/0/0/0, 
    
\texttt{mask=0,}
    
\texttt{psf\_redosky=TRUE,}
    
\texttt{boxiters=2}

\noindent
\texttt{)}

\label{lastpage}
\end{document}